\UseRawInputEncoding
\documentclass[a4paper,fleqn]{cas-sc}

\usepackage[authoryear]{natbib}
\usepackage{caption}
\usepackage{multirow}
\usepackage{colortbl}
\usepackage{enumitem}
\usepackage{threeparttable}
\usepackage{tabularx} 
\usepackage{longtable}
\usepackage{graphicx}
\usepackage{float}
\usepackage{placeins}
\usepackage{mdframed}
\usepackage{booktabs}
\usepackage{listings}
\newfloat{listing}{htbp}{lol}
\floatname{listing}{Listing}

\newdefinition{RQ}{RQ}

\DeclareCaptionStyle{ruled}{labelfont=normalfont,labelsep=colon,strut=off}
\newcolumntype{Y}[1]{>{\raggedright\arraybackslash}p{#1}}

\begin{document}
\let\WriteBookmarks\relax
\renewcommand{\topfraction}{0.95}
\renewcommand{\bottomfraction}{0.95}
\renewcommand{\textfraction}{0.05}
\renewcommand{\floatpagefraction}{0.9}

\shorttitle{}
\shortauthors{K. Qian et al.}

\title[mode = title]{DeanLLM: A Framework for Automated Quality Review of AI-generated Feedback}



\author[1]{Keyang Qian}[type=editor,
                        orcid=0000-0001-7118-3121]
\ead{keyang.qian@monash.edu}

\author[1]{Yixin Cheng}
\ead{yixin.cheng@monash.edu}

\author[1]{Rui Guan}
\ead{rui.guan@monash.edu}

\author[4]{Wei Dai}
\ead{daiw@hku.hk}

\author[5]{Flora Ji-Yoon Jin}
\ead{flora.jin@adelaide.edu.au}

\author[1]{Kaixun Yang}
\ead{kaixun.yang1@monash.edu}

\author[1]{Sadia Nawaz}
\ead{sadia.nawaz@monash.edu}

\author[1]{Zachari Swiecki}
\ead{zach.swiecki@monash.edu}

\author[1]{Guanliang Chen}
\ead{guanliang.chen@monash.edu}

\author[2]{Lixiang Yan}
\cormark[1]
\ead{lixiangyan@mail.tsinghua.edu.cn}

\author[3,1]{Dragan Ga\v{s}evi\'{c}}
\cormark[1]
\ead{dgasevic@hku.hk}

\affiliation[1]{organization={Centre for Learning Analytics, Faculty of Information Technology, Monash University},
    city={Melbourne},
    state={VIC},
    country={Australia}}

\affiliation[2]{organization={School of Education, Tsinghua University},
    city={Beijing},
    country={China}}

\affiliation[3]{organization={Faculty of Education and School of Computing \& Data Science, The University of Hong Kong},
    city={Hong Kong},
    state={Hong Kong SAR},
    country={China}}

\affiliation[4]{organization={Faculty of Education, The University of Hong Kong},
    city={Hong Kong},
    state={Hong Kong SAR},
    country={China}}

\affiliation[5]{organization={College of Education, Behavioural and Social Sciences, Adelaide University},
    city={Adelaide},
    state={SA},
    country={Australia}}

\cortext[cor1]{Corresponding author}

\begin{abstract}
Large language model (LLM) tutors are increasingly used to generate educational feedback, but existing research has focused mainly on feedback generation rather than feedback evaluation. As a result, LLM-generated feedback may offer limited pedagogical value and carry risks of hallucination. The current study introduces \textit{DeanLLM}, an automated review framework for comprehensively evaluating feedback generated by LLM tutors before it is shared with students. We developed a 16-dimension evaluation framework covering feedback content, educational effectiveness, and hallucination risks, and validated it using using human-expert annotations of LLM-generated tutor feedback on synthetic computer science assignment submissions derived from real coursework. We then examined whether LLMs could serve as automated LLM-generated tutor feedback reviewers, and used the best-performing reviewer to benchmark tutor feedback generated by 10 commercial LLMs. Psychometric analyses supported the reliability of the proposed framework and showed that human reviewers tended to evaluate feedback holistically, whereas the LLM reviewer separated rubric dimensions more mechanically. Standard zero-shot and few-shot prompting showed limited agreement with human experts for content-quality judgments. Supervised fine-tuning of GPT-4.1 with human-labelled examples containing scores only, without explanatory rationales, achieved the strongest alignment with expert judgments, reaching 79.8\% accuracy, $\kappa$ = .686, and F1 = 79.4\%. Reasoning LLMs were particularly effective at hallucination detection and produced automated tutor feedback with stronger educational effectiveness and factuality than lightweight models. The findings indicate that \textit{DeanLLM} offers a scalable way for automatically improving the reliability and safety of LLM tutor feedback, while also demonstrating that reviewer calibration and model choice remain critical for educational deployment.
\end{abstract}


\maketitle

\section{Introduction}


Educational feedback is central to learning because it helps students understand their current performance and identify ways to improve future work \citep{h:4,s:4,n:3}. Providing timely, personalised, and high-quality feedback is difficult at scale because feedback requires disciplinary judgement, attention to task goals, and sensitivity to learners’ needs \citep{m:3,r:3,w:6}. Automated feedback systems have long been developed to address this challenge in online learning environments, including but not limited to computing education \citep{c:3,c:1,o:1}. Recent advances in Large Language Models (LLMs) have extended this line of work by enabling natural-language feedback generation across assignments, programming tasks, help requests, and tutor-style interactions \citep{d:1,b:1,h:1,l:1,l:2,p:1,p:2,p:3}.

LLM-generated educational feedback introduces risks that are not fully addressed by existing automated feedback approaches. Researchers have raised concerns about hallucinations in LLM-generated content, including responses that are irrelevant, fabricated, or inconsistent \citep{y:4,l:9}. Some researchers argue that hallucinations are difficult to avoid in LLM generation \citep{x:1,b:5}, while prior work also suggests that feedback can lack relevance or educational value even when it is not strictly hallucinated \citep{z:7}. These risks make it necessary to evaluate both the educational quality and factual reliability of LLM-generated feedback before it reaches students.

Despite extensive research on feedback generation by LLM-based tutors \citep{k:1,d:1,d:2}, limited research has focused on automated evaluation of the quality of feedback generated by LLM tutors based on the feedback text itself. \citet{l:6} and \citet{s:3} utilised LLMs to evaluate LLM-generated feedback for single dimensions, specifically praise components and correctness. \citet{o:1} employed four binary machine learning classifiers to assess feedback effectiveness across task, process, self-regulation, and self levels. \citet{d:4} and \citet{d:5} expanded evaluation frameworks to include readability and content-related dimensions, leveraging LLMs for assessment. However, these studies did not integrate hallucination-related dimensions \citep{z:2,r:2}, despite their importance for reliability, safety, and trustworthiness in educational feedback \citep{a:1}. Prior work suggests that LLMs can support hallucination detection, although their reliability depends on the task, context, and evaluation setup \citep{z:3,r:2,k:2,d:6}.

The present study addressed the lack of automated, multi-dimensional evaluation methods for LLM-generated educational feedback by proposing DeanLLM, an automated review framework for evaluating feedback generated by LLM tutors before delivery to students. DeanLLM applies a 16-dimension feedback evaluation framework that combines feedback content, educational effectiveness, and hallucination-related risks. Based on these evaluations, it identifies low-quality or hallucinated feedback, prompts feedback regeneration by the LLM tutor, and select the most appropriate feedback from multiple candidate options. The study evaluated this feedback evaluation framework by examining the psychometric properties of the proposed 16 dimensions, comparing human expert and LLM-based reviewer evaluations, and exploring prompting and fine-tuning methods for calibrating automated reviewers. The study also assessed the risk that DeanLLM would fail to detect problematic feedback by comparing DeanLLM’s screening decisions with human-labelled evaluations, including false-negative cases in which hallucinated or low-quality feedback is not detected.

Specifically, the contributions of the current study are threefold. First, DeanLLM was introduced as an automated review architecture accompanied by a comprehensive evaluation framework for assessing LLM-generated educational feedback across content quality, pedagogical value, and hallucination risk. Second, prompting and fine-tuning methods were empirically compared to determine how LLMs could be calibrated as automated feedback reviewers aligned with human expert judgments. Third, feedback generated by different LLM tutors was benchmarked using the calibrated automated reviewer, showing how model choice could shape educational quality, factuality, and deployment risks in AI-driven tutoring systems.

\FloatBarrier

\section{Background}

\subsection{Learner-Centered Feedback}
\label{sec:learner-centred-feedback}

Feedback is increasingly conceptualised as a learner-centred process through which students interpret information about their performance and use it to improve subsequent work \citep{b:7,r:3,w:6}. Earlier accounts commonly defined feedback as information that helps learners reduce the gap between current and desired performance \citep{h:4}. More recent learner-centred accounts emphasise students’ active role in making sense of feedback, monitoring their learning, and using feedback to guide future performance \citep{r:3}. This conceptual shift is important for AI-generated feedback because feedback quality cannot be judged only by fluency or correctness. Educational feedback also needs to help learners understand goals, evaluate their current performance, and identify productive next steps.

A learner-centred account of feedback can draw on Hattie and Timperley’s (\citeyear{h:4}) model to specify the direction and focus of useful feedback. The model distinguishes feedback that clarifies goals and standards, feedback that responds to current performance, and feedback that guides future action \citep{h:4,w:8}. The model also distinguishes feedback focused on the task, the processes used to complete the task, learners’ self-regulation, and the self \citep{h:4}. These distinctions are useful for evaluating AI-generated feedback because feedback that supports process understanding, future action, and self-regulation has a stronger pedagogical purpose than general praise or comments directed mainly at the person \citep{h:4,n:3,s:4}.

Effective feedback also depends on the quality of the feedback artefact. Learner-centred feedback should be aligned with learning goals and assessment criteria, specific enough to guide improvement, clear enough for learners to understand, and balanced in its attention to strengths and areas for improvement \citep{b:2,n:3,r:3,s:4}. Feedback that acknowledges strengths can support confidence and engagement, while feedback that identifies weaknesses can help learners develop evaluative judgement and improve future work  \citep{b:2,r:3}. Actionable comments are especially important because they connect feedback on current performance with strategies for subsequent learning  \citep{r:3,s:4}.

Relational qualities further shape whether learners are likely to engage with feedback. Relational feedback recognises that feedback can affect students’ emotions, trust, motivation, and receptivity to critique \citep{h:7, m:3,y:6}. Feedback that is encouraging, respectful, and supportive can strengthen student–instructor relationships and make learners more willing to seek clarification and act on feedback \citep{d:8, m:3,y:6}. These perspectives suggest that evaluating AI-generated educational feedback requires attention to content quality, educational effectiveness, and the relational conditions that make feedback usable for learners.

\subsection{AI-Generated Educational Feedback}
\label{sec:AI-gen-feedback}

Effective feedback can help students understand learning goals, interpret the quality of their current work, and identify productive next steps for improvement \citep{h:4,r:3,w:6}. However, consistently providing such feedback is challenging because it requires substantial instructor time, disciplinary judgement, and sensitivity to learner needs. Earlier automated feedback systems addressed this scaling problem through rule-based, machine-learning, and natural-language-processing approaches, particularly in online learning and programming education \citep{c:3,c:1,o:1}. Recent developments in LLMs have extended this line of work by enabling systems to generate natural-language feedback across a wider range of assignments, code tasks, help requests, and dialogue-based learning activities \citep{d:1,b:1,h:1,l:1,l:2,p:1,p:2,p:3}. In AI tutor contexts, LLM-generated feedback becomes especially consequential because feedback may be shared directly with students as part of an adaptive instructional interaction, which makes pre-delivery review of feedback quality an important condition for responsible use \citep{banihashem2026pedagogical,stamper2024enhancing,cohn2026theory,k:5}.

Research on LLM-generated educational feedback has examined several forms of feedback generation. Studies in computing education have used LLMs to provide feedback on programming assessments, beginner programmers’ help requests, error messages, and code-related tasks \citep{b:1,h:1,l:1,l:2,p:1,p:2,p:3}. Other studies have evaluated LLM-generated feedback in broader assignment and higher education contexts, including comparisons with human or peer feedback sources \citep{d:1,w:7}. Recent evidence also suggests that AI-generated feedback can produce learning outcomes comparable to human feedback in some contexts, although effects vary across feedback quality, learner perceptions, and learning dispositions \citep{k:5}. LLM feedback generation has also been discussed in relation to intelligent tutoring systems and pedagogical agents, where feedback needs to be theoretically grounded, adaptive to learner needs, and evaluated for its educational value rather than treated as a generic text-generation task \citep{stamper2024enhancing,cohn2026theory}. These lines of work position LLM tutors as a relevant context for the present study because tutor-generated feedback can be personalised, interactive, and scalable, but it can also be delivered directly to learners without sufficient quality assurance.

LLM-generated feedback remains vulnerable to pedagogical weaknesses even when the generated text is fluent and grammatically well formed. Feedback may be poorly aligned with task goals, insufficiently specific, unclear to learners, weak in actionable guidance, inappropriate in tone, or limited in its support for future improvement \citep{h:4,s:4,r:3,d:1,d:5}. Empirical studies suggest that learners may value LLM-generated feedback for its availability, clarity, and efficiency, while still raising concerns about correctness, credibility, usefulness, and developmental value \citep{k:4,w:7,alsaiari2025emotionally,pozdniakov2026ai}. These concerns are especially important for AI tutor systems because feedback is often presented as instructional guidance rather than as a draft to be interpreted by a teacher. The educational value of LLM-generated feedback therefore depends on whether feedback satisfies learner-centred criteria for content quality, educational effectiveness, and relational appropriateness before it is delivered to students.

\subsection{Hallucination and Safety Risks in LLM-Generated Feedback}
\label{sec:hallucinations-risks}

Hallucination risks add a further quality-assurance requirement for LLM-generated educational feedback. LLMs can generate responses that appear coherent and authoritative while containing claims that are unsupported, inconsistent with the prompt, or factually incorrect \citep{z:2,r:2,k:2,farquhar2024detecting}. General hallucination research has distinguished errors that arise from contradictions with source input, contradictions within the generated response, and contradictions with external facts \citep{z:2,r:2}. These distinctions are directly relevant to educational feedback. Input-conflicting feedback may misread a student submission or assignment requirement, context-conflicting feedback may offer internally inconsistent advice, and fact-conflicting feedback may present incorrect disciplinary information. Any of these errors can mislead students, reduce trust in the tutor, and weaken the pedagogical value of otherwise well-phrased feedback \citep{y:4,a:1,q:1}.

Educational applications of LLMs require particular caution because hallucinated feedback can shape students’ understanding of their performance and future actions. Recent work on GenAI-driven scaffolding shows that hallucinations in personalised educational support can compromise the learning experience and that pre-delivery evaluation can reduce hallucination risks while improving generated scaffold quality \citep{q:1}. Hallucination detection and mitigation methods in the broader LLM literature provide useful technical foundations, including verification-based prompting, hallucination surveys, and uncertainty-based detection methods \citep{d:6,z:2,r:2,farquhar2024detecting}. These methods also show that hallucination detection remains dependent on task, context, model, and evaluation design. Pedagogical and factual risks therefore make systematic feedback evaluation a necessary complement to LLM-based feedback generation.

\subsection{Automated Feedback Quality Evaluation}
\label{sec:automated-feedback-evaluation}

Feedback evaluation research has used both human-expert rubrics and automated methods to judge the quality of educational feedback. Human evaluation frameworks commonly assess whether feedback is aligned with learning goals, specific, clear, appropriately toned, balanced in its attention to strengths and weaknesses, and actionable for future improvement \citep{h:4,n:3,s:4,b:2,r:3,aldino2025analytics}. Human coding schemes have also been used to evaluate the direction and level of feedback, including feed up, feed back, feed forward, and feedback directed at the task, process, self-regulation, or self \citep{h:4,c:1,o:1,lin2023learner,aldino2025analytics}. Automated approaches have extended this work by using machine learning, natural language processing, and LLM-based methods to classify selected feedback features, such as feedback levels, readability, praise or improvement-oriented components, correctness, and learner-centred, relational, or content-related qualities \citep{o:1,l:6,s:3,d:4,d:5,lin2023learner,aldino2026enhancing,gyamfi2024impact}. These studies show that feedback evaluation can be partially automated, but most existing approaches evaluate selected dimensions rather than a full set of pedagogical qualities needed to judge learner-centred feedback.

Existing feedback evaluation methods provide useful foundations, but they do not yet offer a comprehensive automated approach for reviewing feedback generated by LLM tutors before students receive it. Prior work on educational feedback evaluation has examined content quality and educational effectiveness, while general LLM research has developed taxonomies and detection methods for hallucinations \citep{z:2,r:2,d:6,farquhar2024detecting}. The remaining gap is that hallucination detection has not been sufficiently integrated into automated evaluation of LLM-generated educational feedback alongside pedagogical dimensions such as alignment, specificity, clarity, tone, actionable guidance, feedback direction, and feedback level. A comprehensive review framework is therefore needed to evaluate LLM tutor feedback across feedback content, educational effectiveness, and hallucination risks before delivery to students.

\subsection{Proposed Framework and Research Questions}
\label{sec:research-questions}

Building on the preceding review, the present study addressed the challenge of comprehensively reviewing feedback generated by LLMs before it is delivered to students. Learner-centred feedback research shows that feedback quality depends on multiple pedagogical qualities, including alignment with learning goals, specificity, clarity, tone, actionable guidance, and support for future improvement \citep{h:4,d:4,d:5}. Research on LLM-generated feedback adds a further requirement because feedback may also contain factual, contextual, or input-related errors that can mislead learners \citep{z:2,r:2,a:1}. Existing feedback evaluation approaches therefore need to be extended to assess both the pedagogical value and factual reliability of LLM-generated feedback \citep{o:1,l:6,s:3,d:4,d:5}. The need to evaluate both pedagogical value and factual reliability motivated the development of DeanLLM, an automated review framework for evaluating LLM-generated educational feedback before delivery to students. DeanLLM consists of three linked components. Its evaluation framework defined feedback quality across pedagogical and safety-related dimensions. Its automated reviewer was calibrated to apply the framework in ways that aligned with human expert evaluations. Its benchmarking component used the calibrated reviewer to compare feedback generated by different commercial LLMs.  An overview of the DeanLLM architecture is illustrated in Fig.~\ref{fig:overview2} part a.

The first aim of the study was to establish the evaluation basis of DeanLLM. The study developed and validated an evaluation framework for reviewing LLM-generated educational feedback. The framework operationalised feedback quality through dimensions covering feedback content, educational effectiveness, and hallucination risks. The study therefore first examined whether these dimensions provide a reliable and valid basis for evaluating LLM-generated educational feedback, and how LLM-based evaluations compare with human expert evaluations.

\begin{RQ}
To what extent does the proposed evaluation framework provide a reliable and valid basis for evaluating LLM-generated educational feedback, and how do LLM-based evaluations compare with human expert evaluations?
\end{RQ}
The second aim of the study was to calibrate the automated review component of DeanLLM \citep{d:5,h:5,z:6}. Different prompting and fine-tuning approaches may produce different levels of agreement with human expert evaluations, making reviewer calibration central to the framework \citep{d:5,l:10,l:12}.

\begin{RQ}
Which prompting or fine-tuning approach produces the strongest alignment between automated feedback reviewers and human expert evaluations?
\end{RQ}
The third aim of the study was to examine how DeanLLM can be used to evaluate feedback generated by different LLMs \citep{d:1,b:1,p:1,p:3}. Model choice may affect the pedagogical quality and factual reliability of generated feedback, which is important for understanding the conditions under which LLM-generated feedback can be used responsibly \citep{d:1,q:1,y:4}.

\begin{RQ}
How does the pedagogical quality and factual reliability of feedback vary across commercial LLMs when they are used as tutors to generate educational feedback?
\end{RQ}
\section{Method}
\label{sec:method}

\subsection{Study Design, Data Source, and Ethics}
\begin{figure}[pos=htbp]
  \centering
    \includegraphics[width=\textwidth]{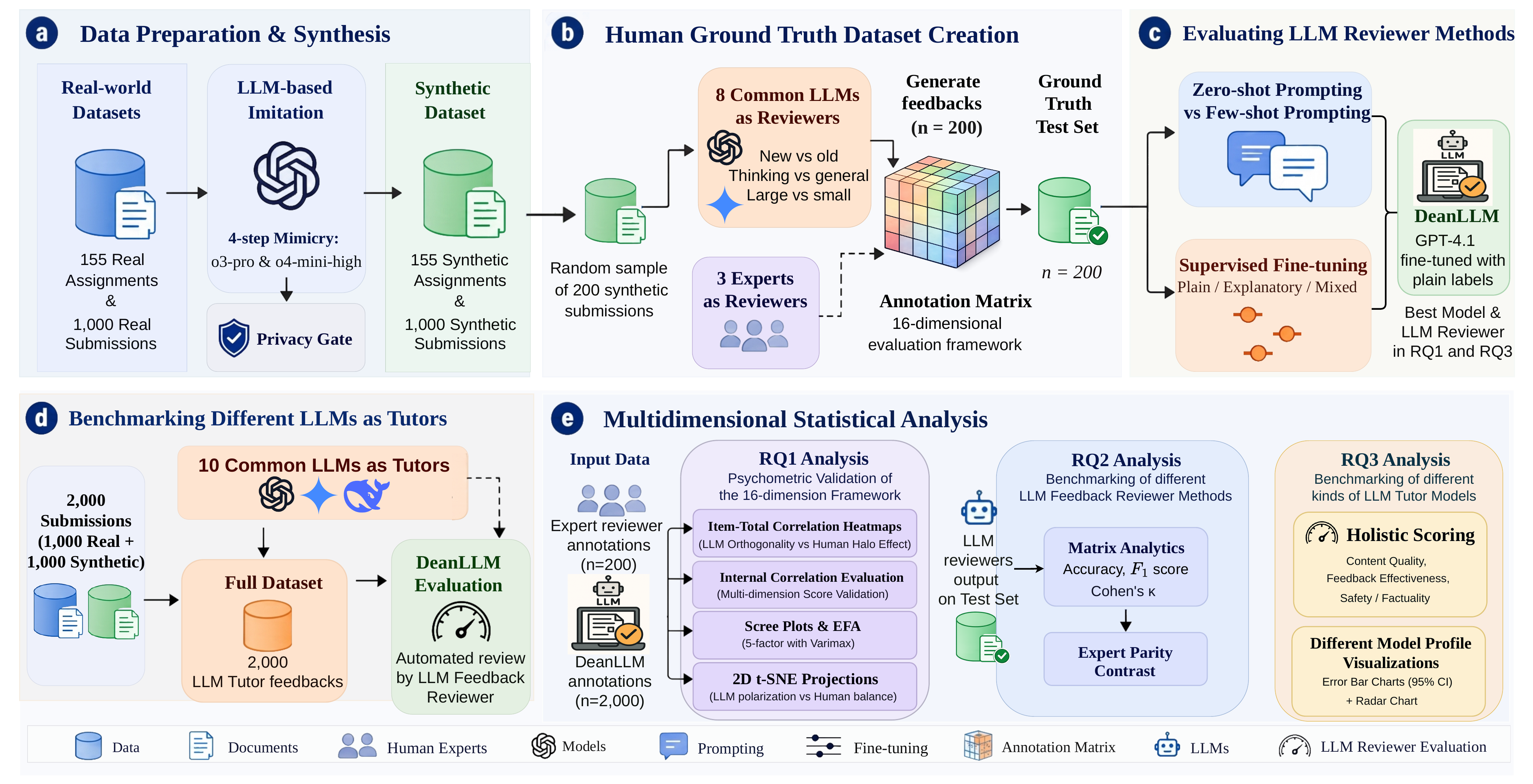}
      \caption{Overview of experiment process and data flow. The study comprises five integrated stages: (a) data preparation and synthesis; (b) human ground truth dataset creation; (c) evaluating different LLM feedback reviewer methods; (d) benchmarking different types of LLMs as tutors; and (e) multi-dimensional statistical analysis across three RQs.}
      \label{fig:overview}
\end{figure}

The study used a three-stage offline design to develop, calibrate, and benchmark DeanLLM as an automated reviewer of LLM-generated educational feedback. Figure~\ref{fig:overview} summarises the data preparation (a), human labelling (b), reviewer calibration (c), benchmarking of different LLM tutors (d), and multi-dimensional analyses that connected the three research questions (e). The first stage defined and evaluated a 16-dimension evaluation framework and produced a human-labelled ground truth dataset for feedback quality evaluation (RQ1). The second stage compared prompting and fine-tuning approaches for implementing LLM feedback reviewers against the human-labelled dataset (RQ2). The third stage used the best-performing reviewer configuration to evaluate feedback generated by 10 commercial LLM tutors on paired real and synthetic computer science assignment submissions (RQ3).

\begin{figure}[pos=htbp]
  \centering
    \includegraphics[width=\textwidth]{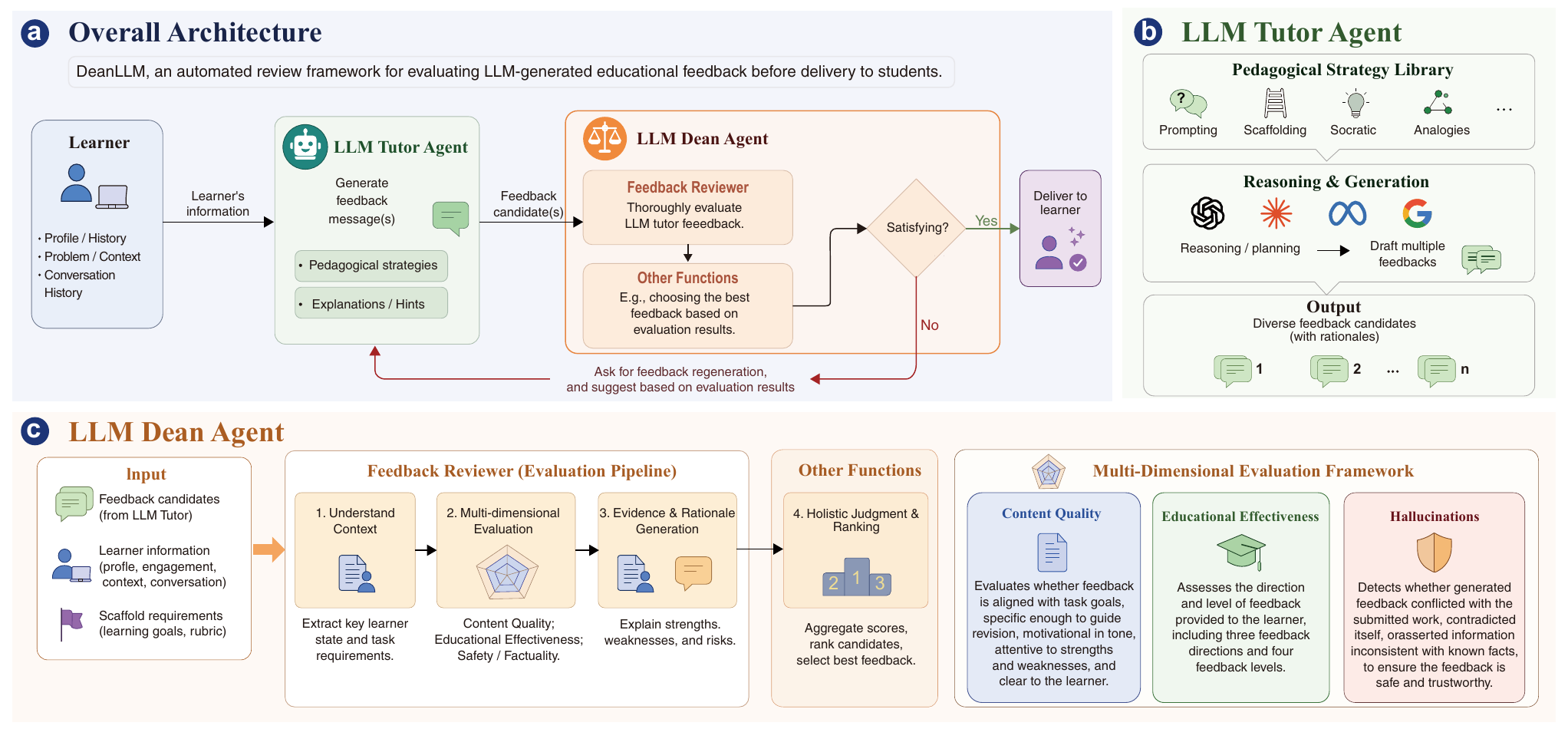}
      \caption{Overview of the DeanLLM architecture, an automated review framework designed to evaluate LLM-generated educational feedback before it is delivered to students. The workflow illustrates the interaction between the learner, the LLM Tutor Agent, and the LLM Dean Agent. The LLM Tutor Agent utilizes learner information and pedagogical strategies to generate feedback candidate(s). The LLM Dean Agent then acts as an automated feedback reviewer, assessing feedback candidate(s) through a multi-dimensional evaluation framework that measures Content Quality, Educational Effectiveness, and Hallucinations, which is the focus of our work. Depending on this evaluation, the Dean Agent can choose the most appropriate feedback candidate, and then either approves for delivery to the learner or triggers a regeneration request to ensure pedagogical quality of the feedback.}
      \label{fig:overview2}
\end{figure}

The source coursework data were retrieved from university computer science courses delivered during the first semester of 2021 at a [country blinded] university (Fig.~\ref{fig:overview} part a). The retrieval was approved by the [country blinded] University Human Research Ethics Committee (application no. [blinded]), and no new student participants were recruited for the present study. The source materials included assignment descriptions, rubrics, supplementary assignment materials, and student submissions. These source materials were converted into text-readable formats for LLM processing, tables were represented as structured text, and figures were excluded from the prompt inputs. Details of the real data retrieval process are provided in Appendix~\ref{appx:data_retrieval}.

The analytic coursework sample consisted of paired real and synthetic assignment records drawn from the eligible source dataset. The real assignment data could not be publicly released because they contained student information and institution-owned coursework material. To support reproducible evaluation while protecting privacy and copyright, a one-to-one LLM-based imitation procedure was used to generate 1,000 synthetic assignment submissions and corresponding synthetic assignment descriptions spanning 155 assignments from 59 courses. Each synthetic assignment description was generated from a real assignment description, and each synthetic submission was generated from the corresponding real submission and paired real--synthetic assignment descriptions.
\label{sec:method_synthesis}
The synthetic data generation procedure was designed to preserve task structure without copying private or copyrighted content. The assignment-description synthesis used the o3-pro LLM with high reasoning effort to imitate the theme, objectives, writing style, length, and rubric detail of each real assignment description. The submission synthesis used the o4-mini LLM with high reasoning effort to imitate the level of detail, correctness, writing style, and approximate length of each real student submission in response to the synthetic assignment description. The synthesis prompts followed a four-step procedure which included: evaluation of the source record, generation of a synthetic counterpart, evaluation of the synthetic output on the same dimensions, and revision when the synthetic output did not sufficiently match the source record. The full prompt formats are reported in Appendices~\ref{appx:description_prompt} and~\ref{appx:submission_prompt}.

The synthetic data pipeline included validity and privacy checks before the synthetic records were used for feedback generation. Semantic resemblance was assessed with BERTScore F1, and length preservation was assessed with Pearson correlations between real and synthetic text lengths, as reported in Appendix~\ref{appx:synthesis}. A separate privacy gate used o4-mini with high reasoning effort to compare names and student identifiers detected in the real and synthetic submissions. Synthetic submissions were rejected and regenerated when the privacy gate indicated that an identifier from the real submission had been preserved. All model calls for data synthesis were made through API endpoints configured so that submitted data were not used for future model training.

\subsection{Feedback Evaluation Framework}
\label{sec:framework}

The feedback evaluation framework operationalised LLM-generated educational feedback quality across content, educational effectiveness, and hallucination risk, as illustrated in the bottom right of Fig.~\ref{fig:overview2} part c. The framework contained 16 dimensions: six dimensions assessed feedback content on ordinal 0--2 scales, seven dimensions assessed educational effectiveness as binary indicators, and three dimensions assessed hallucination types as binary indicators. The dimensions were selected from learner-centred feedback \citep{h:4,r:3,w:6,n:3,s:4}, automated feedback evaluation \citep{o:1,l:6,s:3,d:4,d:5}, and LLM hallucination literatures \citep{l:9,z:2,r:2,a:1} to capture both pedagogical value and factual reliability before feedback is delivered to students. Detailed annotation rubric for each dimension is open-sourced (see Sec.~\ref{sec:opensource}).

Content-quality dimensions captured whether the feedback was aligned with task goals, specific enough to guide revision, motivational in tone, attentive to strengths and weaknesses, and clear to the learner \citep{h:4,b:2,n:3,r:3,s:4,h:7,m:3,y:6}. Each content-quality dimension was rated on a three-point scale, where 0 indicated absence or inadequacy of the dimension, 1 indicated partial or vague presence, and 2 indicated clear and consistent presence. Table~\ref{tab:feedback_content_dimensions} defines the six content-quality dimensions used in the coding rubric. 

\begin{table}[pos=htbp]
\centering
\caption{Feedback content dimensions graded using a 3-point scale.}
\label{tab:feedback_content_dimensions}
\begin{tabular}{@{}Y{0.20\linewidth}@{}Y{0.60\linewidth}@{}Y{0.16\linewidth}@{}}
\toprule
\textbf{Dimension} & \textbf{Operational definition} & \textbf{Source} \\
\midrule
Alignment with goals & Assesses whether most feedback comments are aligned with specific learning goals, task requirements, or rubric criteria. & \citep{n:3} \\
Specificity & Assesses whether feedback offers detailed, actionable guidance and includes concrete examples or directions for improvement. & \citep{s:4,g:2} \\
Motivational Tone & Assesses whether the feedback is consistently positive, encouraging, respectful, and appropriate for learner support. & \citep{n:3,l:6,d:5} \\
Strength & Assesses whether the feedback identifies specific strengths in the student submission. & \citep{r:3,b:2,d:4} \\
Weakness & Assesses whether the feedback identifies specific weaknesses or areas requiring revision in the student submission. & \citep{r:3,b:2,d:4} \\
Clarity and Understandability & Assesses whether the feedback language is clear, comprehensible, and easy for the learner to interpret. & \citep{r:3,b:2,d:4} \\
\bottomrule
\end{tabular}
\end{table}
\FloatBarrier

Educational-effectiveness dimensions captured the direction and level of feedback provided to the learner. The first three dimensions distinguished feed up, feed back, and feed forward, corresponding to feedback about goals, current progress, and next steps. The remaining four dimensions distinguished task, process, self-regulation, and self-level feedback. Each dimension was coded as 1 when the feature was present in the feedback and 0 when it was absent. Table~\ref{tab:feedback_effectiveness_dimensions} defines the seven educational-effectiveness dimensions used in the coding rubric. 

\begin{table}[pos=htbp]
\centering
\caption{Feedback effectiveness dimensions labelled as binary indicators.}
\label{tab:feedback_effectiveness_dimensions}
\begin{tabular}{@{}Y{0.20\linewidth}@{}Y{0.60\linewidth}@{}Y{0.16\linewidth}@{}}
\toprule
\textbf{Dimension} & \textbf{Operational definition} & \textbf{Source} \\
\midrule
\multicolumn{3}{@{}l}{\textit{Feedback direction}} \\
\midrule
Feed forward & Indicates feedback about next steps in learning, future tasks, or actions the learner can take to improve subsequent work. & \citep{h:4,d:4,d:5} \\
\cmidrule{1-3}
Feed up & Indicates feedback that clarifies the learning goal, task goal, or criteria for successful performance. & \citep{h:4,d:4,d:5} \\
\cmidrule{1-3}
Feed back & Indicates feedback about the learner's current progress, performance, or response to the submitted work. & \citep{h:4,d:4,d:5} \\
\midrule
\multicolumn{3}{@{}l}{\textit{Feedback level}} \\
\midrule
Feedback on task & Indicates feedback about how well the task was performed, including correctness, missing information, and surface-level task completion. & \citep{h:4,c:1} \\
\cmidrule{1-3}
Feedback on process & Indicates developmental guidance about strategies, procedures, reasoning processes, information processing, or ways of completing future work. & \citep{d:7,c:1} \\
\cmidrule{1-3}
\shortstack[l]{Feedback on\\ self-regulation} & Indicates feedback that supports monitoring, self-assessment, confidence, planning, or regulation of future learning. & \citep{d:7,c:1} \\
\cmidrule{1-3}
Feedback on self & Indicates personal praise or comments directed at the learner as a person rather than at the task, process, or regulation of learning. & \citep{d:7} \\
\bottomrule
\end{tabular}
\end{table}
\FloatBarrier

Hallucination dimensions were defined in line with prior accounts of LLM hallucinations as fabricated, irrelevant, or inconsistent content \citep{l:9,z:2,r:2} and with the hallucination-risk background reviewed in Section~\ref{sec:hallucinations-risks}. These dimensions captured whether generated feedback conflicted with the submitted work, contradicted itself, or asserted information inconsistent with known facts, to ensure the feedback is safe
and trustworthy. Each hallucination dimension was coded as 1 when the hallucination type was present and 0 when it was absent. Table~\ref{tab:types_of_hallucinations} defines the three hallucination dimensions used in the coding rubric.

\begin{table}[pos=htbp]
\centering
\caption{Hallucination dimensions labelled as binary indicators.}
\label{tab:types_of_hallucinations}
\begin{tabular}{@{}Y{0.20\linewidth}@{}Y{0.60\linewidth}@{}Y{0.16\linewidth}@{}}
\toprule
\textbf{Hallucination type} & \textbf{Operational definition} & \textbf{Source} \\
\midrule
Input-conflicting Hallucination & Indicates feedback that replaces, misreads, or contradicts information contained in the assignment description or student submission. & \citep{z:2,r:2} \\
Context-conflicting Hallucination & Indicates feedback that contains internally contradictory claims within the same feedback response. & \citep{z:2,r:2} \\
Fact-conflicting Hallucination & Indicates feedback that makes claims contradicting known facts or the factual constraints of the task context. & \citep{z:2,r:2} \\
\bottomrule
\end{tabular}
\end{table}
\FloatBarrier

\subsection{Human-labelled Ground Truth Dataset}
\label{sec:human_labels}

The human-labelled ground truth dataset was constructed from the synthetic coursework records (mentioned in the previous Method Sec.~\ref{sec:method_synthesis}) to support public release while preserving the structure of real assessment tasks (Fig. \ref{fig:overview} part b). A random sample of 200 synthetic assignment submissions and their corresponding synthetic assignment descriptions was selected from the synthetic dataset. The sample spanned 85 assignments from 43 courses. Eight LLM tutors generated feedback for this dataset, with 25 feedback instances generated by each tutor model, producing 200 feedback instances for human coding.

The tutor models used to construct the human-labelled dataset were selected to cover different model families and capability levels. The eight tutor models were OpenAI GPT-4.1 nano, GPT-4.1, o4-mini, and o3, and Google Gemini 2.0 Flash-Lite, Gemini 2.0 Flash, Gemini 2.5 Flash Preview, and Gemini 2.5 Pro Preview. Each model received the assignment description and the synthetic student submission and generated one feedback response. Where reasoning-effort parameters were available, reasoning models were run with high reasoning effort, and the temperature was set to 0 for feedback generation.

Human feedback reviewers were recruited because the calibration dataset required domain-informed judgements about feedback quality and hallucination risk. Three researchers with experience in LLM-generated educational feedback and feedback effectiveness served as coders. The coders first discussed the 16-dimension rubric and then independently labelled 16 calibration feedback instances, with two instances drawn from each of the eight tutor models. Inter-rater reliability on this calibration set reached Fleiss' $\kappa=0.637$. Disagreements in the calibration set were discussed to refine the interpretation of the rubric before the remaining instances were labelled.

The remaining feedback instances were labelled after coder calibration with balanced allocation across tutor models. Each coder received feedback instances generated by each tutor model so that model source was not concentrated within one coder's workload. During hallucination coding, coders could consult non-binding hallucination labels and explanations generated by o3 and o3-pro as auxiliary references, but the human coder retained responsibility for the final label. The human-labelled dataset contained the 16 rubric labels for each feedback instance and formed the ground truth for reviewer calibration and evaluation.

Explanatory annotations were collected to support fine-tuning conditions that used human rationales. Coders were asked to add explanations when a label was difficult to judge, especially for hallucination dimensions. This process produced 45 labelled feedback instances with explanations. The explanatory instances were retained separately from plain-labelled instances so that fine-tuning could compare labels-only supervision with rationale-augmented supervision. An example of a human annotation with explanation is reported in Appendix~\ref{appx:annotation_example}.

\subsection{LLM Tutor Feedback Generation}
\label{sec:tutor_feedback}

In our study, LLM tutor agents were implemented as single-turn generative AI agents that produced formative feedback on assignment submissions (Fig. ~\ref{fig:overview} part d), serving as a specific instantiation of the LLM Tutor Agent component within the DeanLLM architecture (Fig.~\ref{fig:overview2} part b). The benchmarking dataset used 10 commercial LLM tutor backbones: OpenAI o3-2025-04-16, o4-mini-2025-04-16, gpt-4.1-2025-04-14, and gpt-4.1-nano-2025-04-14; Google gemini-2.5-pro, gemini-2.5-flash, gemini-2.0-flash-001, and gemini-2.0-flash-lite-001; and DeepSeek-V3-0324 and DeepSeek-R1-0528. These models were selected to include reasoning models, general-purpose high-capacity models, and lightweight models. Each tutor model generated 100 feedback responses for real assignment submissions and 100 feedback responses for the corresponding synthetic submissions, yielding 2,000 tutor-feedback instances.

The tutor role was specified behaviourally rather than as an anthropomorphic persona. The prompt instructed the model to act as a virtual tutor and generate educational feedback for the provided student submission in relation to the provided assignment description. The prompt disclosed the task role inside the model instruction, but no feedback was delivered to students and no live student interaction occurred. The prompt also instructed the model not to penalise students for missing figures, excluded files, or table conversions caused by preprocessing, and it required each feedback response to be fewer than 300 words. The full tutor prompt is reported in Appendix~\ref{appx:tutor_prompt}.

Tutor generation was executed as a non-interactive single-turn procedure. Each tutor received only the assignment description and the corresponding student submission, and each model produced one feedback response for each assigned case. Models did not interact with one another, did not revise outputs after receiving reviewer judgements, and did not participate in multi-agent turn-taking. The prompt input, model identifier, source type of the assignment record, and generated feedback text were logged for subsequent review.

Synthetic and real tutor-feedback outputs were compared before the synthetic and real feedback records were pooled for reviewer evaluation. The comparison used Ahmad et al.'s feedback evaluator \citep{a:2} to code seven learner-centred feedback attributes in feedback generated for paired real and synthetic submissions. Alignment, sensitivity, specificity, and F1 score were computed by treating the attributes of feedback generated for real submissions as the reference labels. The utility comparison is reported in Appendix~\ref{appx:feedback_evaluation}; the pooled tutor-feedback dataset used in subsequent analyses contained 1,000 real-submission feedback instances and 1,000 synthetic-submission feedback instances.

\subsection{LLM Feedback Reviewer Implementation}
\label{sec:reviewers}

LLM feedback reviewers were implemented as automated rubric scorers for LLM tutor feedback, representing an essential instantiation of the LLM Dean agent within the overarching DeanLLM architecture (illustrated in Fig.~\ref{fig:overview2} part c). To evaluate different feedback reviewer methods and select the optimal configuration to serve as the primary DeanLLM for subsequent analyses (Fig.~\ref{fig:overview} part c),  four OpenAI models were used: GPT-4.1, o4-mini, o3, and o3-pro. The reviewer role was to read the assignment description, the student submission, the generated tutor feedback, and the 16-dimension rubric, and then output labels for all feedback-quality and hallucination dimensions. The reviewer role was not concealed from human participants because no student or teacher interacted with the reviewer during the experiment.

Reviewer persona manipulation was limited to task stance, output discipline, and rubric adherence. Reviewers were instructed to behave as feedback reviewers rather than tutors, to evaluate the supplied feedback rather than regenerate it, and to follow the requested output schema. Reviewers received the same substantive task knowledge across conditions: the assignment description, the corresponding student submission, the tutor feedback to be reviewed, and the labelling rubric. The reviewer was not given access to external sources, student identities, or live course context beyond the prompt inputs. 

Reviewer prompting conditions manipulated the amount of rubric support provided at inference time. In the zero-shot condition, the prompt supplied the assignment description, student submission, tutor feedback, rubric definitions, and output-format requirements, but excluded labelled examples. In the few-shot condition, the prompt used the same structure but added labelled examples and explanations for different values of the rubric dimensions. Prompt examples for the Alignment with goals dimension are reported in Appendices~\ref{appx:zero-prompt} and~\ref{appx:few-prompt}.

Reviewer fine-tuning conditions manipulated the supervision used to calibrate GPT-4.1 as a feedback reviewer. The human-labelled dataset was split into training and test partitions with feedback generated by different tutor models distributed evenly across the partitions. The plain-labelled fine-tuning condition used 100 training instances in which the input was the few-shot reviewer prompt and the expected output was the human label vector. The explanatory fine-tuning condition used 45 training instances in which the expected output contained the human labels followed by the coder's explanation. The mixed fine-tuning condition first fine-tuned on a random half of the explanatory instances, then on all plain-labelled instances, and then on the remaining explanatory instances. All GPT-4.1 fine-tuning runs used \texttt{n\_epochs}=2.

Reviewer calls were executed as logged, non-interactive, single-turn evaluations. Each reviewer produced one label vector for each assigned tutor-feedback instance, and no reviewer could ask follow-up questions, revise labels after seeing human labels, or use another model's output during test-time evaluation. Dominance prevention was not relevant because the experiment did not include multi-agent discussion; instead, comparability was maintained by holding the task inputs and rubric constant within each reviewer condition. The reviewer implementation was validated by comparing reviewer outputs with the human-labelled ground truth dataset.

\subsection{Measures and Composite Scores}

The primary coded measures were the 16 rubric dimensions defined in Tables~\ref{tab:feedback_content_dimensions}--\ref{tab:types_of_hallucinations}. Higher scores on the six content-quality dimensions indicated stronger presence of the desired feedback feature. A value of 1 on the seven educational-effectiveness dimensions indicated that the feedback feature was present. A value of 1 on any hallucination dimension indicated that the corresponding risk was present, so hallucination labels were reverse-interpreted when constructing safety scores.

\label{sec:feedback_scores}
The three composite scores summarised feedback quality on a 0--100 scale for model benchmarking. \textit{Content Quality} was computed as $100 \times (\bar{x}/2)$, where $\bar{x}$ was the mean of the six 0--2 content ratings for a feedback instance. For \textit{Educational Effectiveness}, we designed a heuristic weighting mechanism grounded in established pedagogical theory and formative feedback frameworks \citep{h:4, s:4}. Because different feedback levels yield disparate impacts on student learning outcomes, we assigned varying weights to the binary indicators ($F$) detected by the model. Higher weights (0.40) were allocated to features known to trigger deep cognitive and metacognitive growth: feed-forward ($F_{\mathrm{forward}}$), process ($F_{\mathrm{process}}$), and self-regulation ($F_{\mathrm{selfreg}}$). Conversely, surface-level task feedback ($F_{\mathrm{task}}$) was assigned a lower weight (0.20), and feedback on the self ($F_{\mathrm{self}}$) was weighted at 0.00, as meta-analyses demonstrate it offers minimal instructional value and can harm self-efficacy \citep{h:4, s:4}. Accordingly, \textit{Educational Effectiveness} was computed as $100 \times (0.40F_{\mathrm{forward}} + 0.20F_{\mathrm{task}} + 0.40F_{\mathrm{process}} + 0.40F_{\mathrm{selfreg}} + 0.00F_{\mathrm{self}})/1.40$. Feed up and feed back were retained for dimension-level analyses and construct validation but were not included in this weighted utility index. \textit{Safety/Factuality} was scored as 100 when all three hallucination indicators equalled 0 and as 0 when any input-conflicting, context-conflicting, or fact-conflicting hallucination was labelled. The three composite scores were reported separately for model benchmarking; any overall profile treated them as parallel dimensions rather than as evidence of student learning gains. To ensure that these heuristic weighting mechanisms and strict veto rules are mathematically robust and pedagogically justified, their validity was empirically verified against alternative scoring configurations via comprehensive sensitivity analyses (detailed in Appendix~\ref{appx:sensitivity}).

Agreement measures quantified the alignment between automated reviewers and human feedback reviewers. Accuracy represented the proportion of matching labels between a reviewer output and the human-labelled ground truth. F1 summarised classification performance for the coded labels within each construct. Cohen's $\kappa$ measured agreement beyond chance for each reviewer configuration, and Fleiss' $\kappa$ was used for the three-coder calibration set. Human inter-rater agreement from the calibration set was used as a baseline for interpreting automated reviewer agreement.

\subsection{Analysis Plan}
\label{sec:analysis_plan}

RQ1 was answered with psychometric and structural analyses of the 16-dimension evaluation framework (Fig.~\ref{fig:overview} part e). The analysis used the human-labelled dataset ($n=200$) and the DeanLLM-labelled tutor-feedback dataset ($n=2,000$), where DeanLLM refers to the best-performing reviewer configuration selected in RQ2. Pearson's correlation matrices were computed to examine collinearity and independence among rubric dimensions in each dataset. Corrected item-total correlations were computed within the theoretically defined constructs (Feedback content, Feedback effectiveness, and Hallucinations, as discussed in previous Method Sec.~\ref{sec:framework}) to assess whether dimensions aligned with our intended composite scores. Exploratory factor analysis was then used to examine latent structure, with scree plots and the Kaiser criterion inspected before applying a five-factor baseline solution with Varimax rotation for cross-population comparability. Factor loadings were interpreted using a primary threshold of 0.30. t-distributed stochastic neighbour embedding (t-SNE) was used to project dominant factor patterns into two dimensions after standardising the coded dimensions. Robustness checks tested four-, five-, and six-factor extractions, loading thresholds of 0.30, 0.40, and 0.50, and an 80/20 split-sample stability check with random state 999, as detailed in Appendix~\ref{appx:robustness}. This analysis answered whether the framework showed coherent construct behaviour and whether human and LLM-based evaluations organised the same rubric dimensions in similar ways. Finally, to empirically validate the theoretical assumptions underpinning the composite score calculations, sensitivity analyses tested alternative configurations—specifically, an equal-weighting scheme for Educational Effectiveness and a graduated penalty scheme for Safety \& Factuality—to determine whether the chosen heuristics accurately differentiated model capabilities and risk profiles (detailed in Appendix \ref{appx:sensitivity}).

RQ2 was answered with reviewer-agreement analyses against the human-labelled ground truth dataset (Fig. \ref{fig:overview} part e). The focal dependent variables were the 16 rubric labels, summarised overall and within Content Quality, Educational Effectiveness, and Hallucinations. The main grouping variables were reviewer backbone, prompting condition, and fine-tuning condition. Zero-shot and few-shot reviewer configurations were evaluated for GPT-4.1, o4-mini, o3, and o3-pro. Fine-tuned reviewer configurations were evaluated for GPT-4.1 using plain-labelled data, explanatory data, and mixed plain-plus-explanatory data. Accuracy, F1, and Cohen's $\kappa$ were computed for each reviewer configuration on the held-out test partition, and the human coder calibration agreement was used as the baseline. The best-performing reviewer was selected by overall agreement with human labels, with separate attention to hallucination agreement because hallucination errors determine the safety-gating score.


RQ3 was answered with model-level benchmarking of feedback generated by commercial LLM tutors (Fig. \ref{fig:overview} part e). Specifically, we employed 10 representative commercial LLMs to act as tutors: OpenAI (GPT-4.1 nano, GPT-4.1, o4-mini-high, and o3-high), Google (Gemini 2.0 Flate-Lite, Gemini 2.0 Flash, Gemini 2.5 Flash, and Gemini 2.5 Pro), and Deepseek (Deepseek R1 and Deepseek V3). These models were selected to represent a diverse array of current paradigms, including chain-of-thought reasoning models (Gemini 2.5 Pro, o3-high, o4-mini-high, Deepseek R1), general large models (GPT-4.1, Gemini 2.0 Flash, Gemini 2.5 Flash, Deepseek V3), and cost-efficient small models (GPT-4.1 nano, Gemini 2.0 Flate-Lite). Each tutor model generated 100 feedback instances for synthetic assignment submissions and 100 for original real assignments, resulting in a total of 2,000 feedback samples. To evaluate this feedback, the best-performing evaluation model identified in RQ2 (i.e., the GPT-4.1 model fine-tuned with plain-labelled data) served as our \textit{DeanLLM} reviewer. The focal dependent variables extracted by the DeanLLM were the three 0--100 composite scores: Content Quality, Educational Effectiveness, and Safety/Factuality. The main grouping variable was tutor model, and the data source of the feedback prompt (real versus synthetic submission) was retained as a design attribute for checking whether pooled comparisons were affected by source type. Model-level means and 95\% confidence intervals were reported for each composite score, and score differences were interpreted on the 0--100 scale. Hallucination results were also interpretable as differences in the proportion of feedback instances receiving any hallucination label because the Safety/Factuality score used a strict zero-tolerance rule.

All analyses treated feedback instances as the unit of analysis unless a model-level mean was explicitly reported. The analyses did not estimate learning effects, student outcomes, or causal effects of receiving feedback because no feedback was delivered to students. The inferential scope was therefore limited to the reliability of the evaluation framework, the alignment between automated and human feedback review, and the comparative quality of feedback generated by the evaluated LLM tutor models.

\section{Results}
\label{sec:results}

\subsection{RQ1: Psychometric Structure of the LLM Feedback Evaluation Framework}
\label{sec:results_rq1}

\begin{figure}[pos=htbp]
  \centering
  \includegraphics[width=\linewidth]{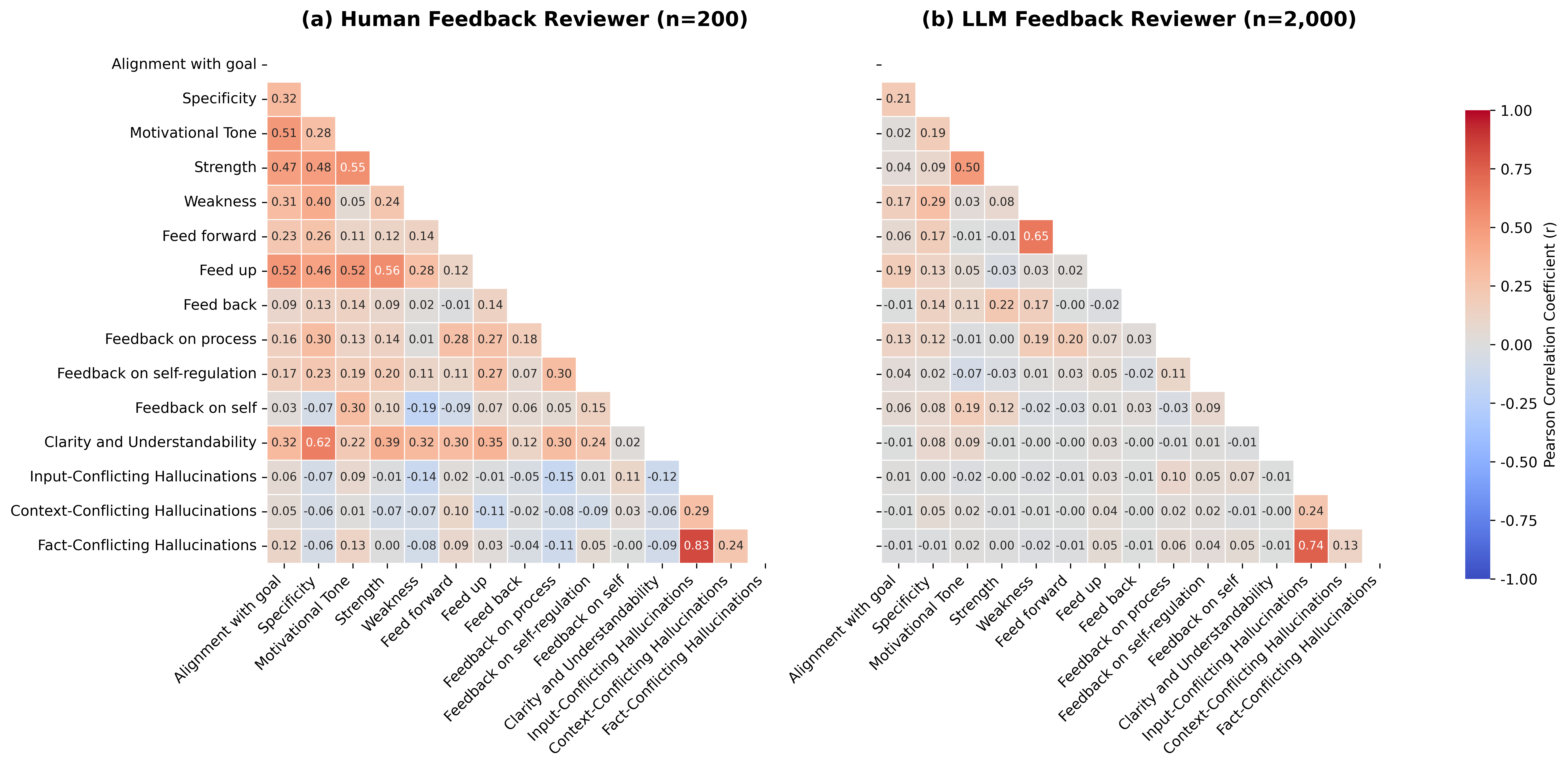}
  \caption{Comparison of inter-dimensional Pearson correlation matrices between (a) human feedback-reviewer evaluation results ($n=200$) and (b) LLM feedback-reviewer evaluation results ($n=2{,}000$).}
  \label{fig:Combined_Feedback_Correlation_Heatmap}
\end{figure}

The inter-dimensional correlation analysis for RQ1 showed different rating structures in human and automated LLM feedback-reviewer annotations. In the human-labelled dataset ($n=200$), several feedback-quality dimensions were positively associated with one another, whereas the LLM-labelled dataset ($n=2{,}000$) showed near-zero correlations for most dimension pairs. Figure~\ref{fig:Combined_Feedback_Correlation_Heatmap} shows that human reviewers tended to produce a more interdependent rating profile, while the LLM feedback reviewer produced a more separated dimension-by-dimension rating profile.

\begin{figure}[pos=htbp]
  \centering
  \includegraphics[width=\linewidth]{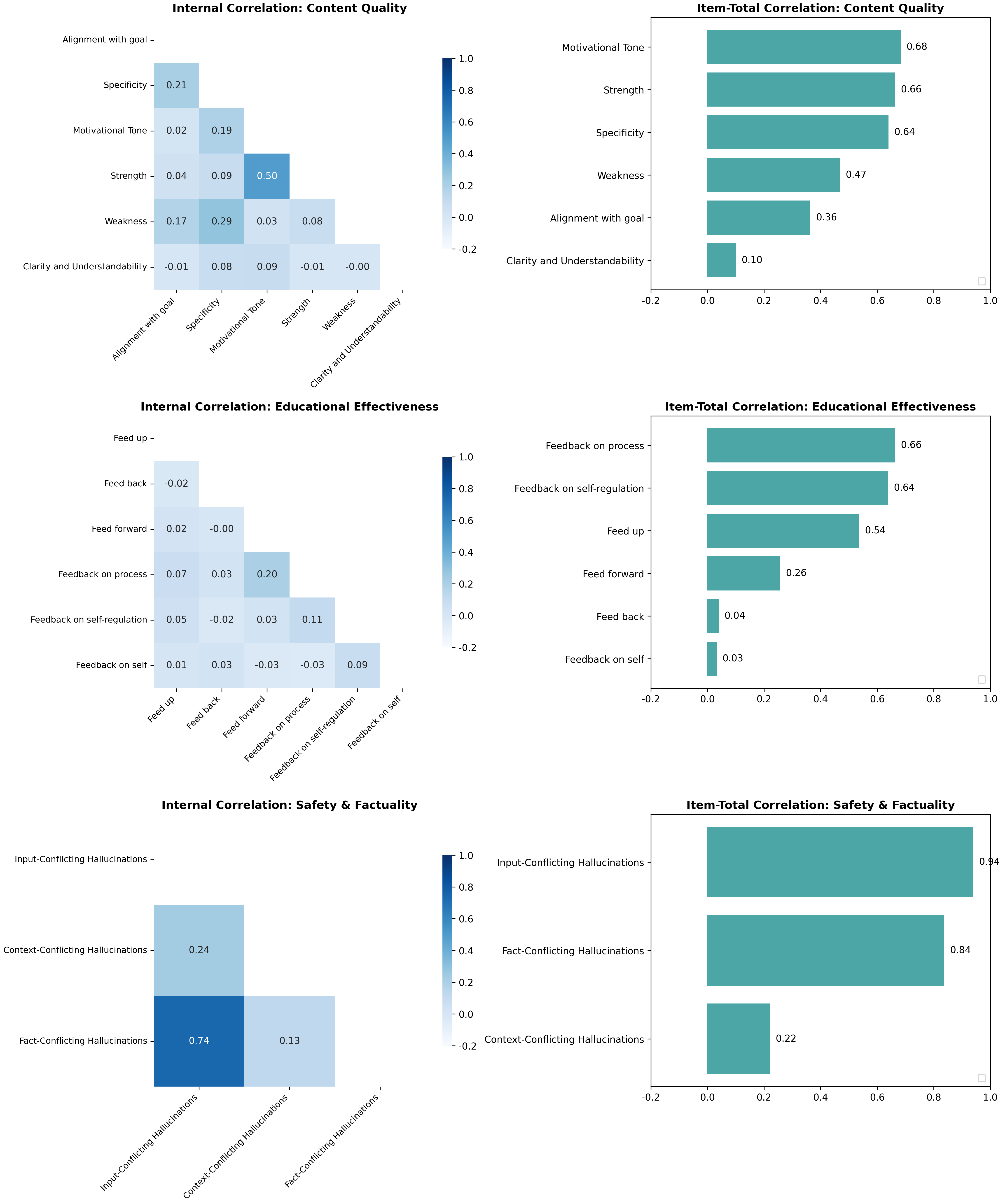}
  \caption{Construct-level evaluation of Content Quality, Educational Effectiveness, and Safety \& Factuality. The left column displays internal correlation matrices for the subdimensions, and the right column reports item-total correlations.}
  \label{fig:scale_construct_validation}
\end{figure}

Construct-level item-total analyses, as illustrated in Fig.~\ref{fig:scale_construct_validation}, showed that the three evaluation constructs had different internal patterns, each supported by theoretical and empirical rationales. For Content Quality, core indicators including Motivational Tone, Strength, Specificity, Weakness, and Alignment with goal demonstrated strong item-total correlations ranging from 0.36 to 0.68, confirming excellent convergent validity. While Clarity and Understandability showed a lower item-total correlation (0.10), this was primarily driven by a ceiling effect: modern LLMs inherently generate highly fluent text, severely compressing the variance of this specific metric. For Educational Effectiveness, the empirical data perfectly aligned with our theory-driven weighting formula in Method Sec.~\ref{sec:feedback_scores}. High-order cognitive feedback dimensions, such as Feedback on process and Feedback on self-regulation---which were assigned the highest weights (0.4) in our formula---yielded the highest item-total correlations (both $r > 0.63$). Conversely, Feedback on self showed a near-zero correlation ($r = 0.03$), statistically justifying its 0.0 weight in our theoretical weight distribution of composite score of the Educational Effectiveness construct. Finally, for Safety \& Factuality, which operates as a formative model using a maximum-penalty mechanism, Input-Conflicting and Fact-Conflicting hallucinations emerged as the primary drivers of the safety score (both $r > 0.83$). Context-Conflicting hallucinations showed a lower correlation ($r = 0.22$) due to the extreme sparsity of such errors in our dataset; nevertheless, its retention remains theoretically mandated by the strict zero-tolerance nature of educational safety evaluation.

Appendix \ref{appx:sensitivity} reports sensitivity analyses on the composite scoring weights, confirmed that the theoretical weighting formula effectively isolates and rewards models capable of higher-order cognitive scaffolding. When evaluated under an equal-weight assumption, reasoning models that provided deep process-level feedback experienced a score reduction, whereas models generating surface-level praise saw artificial score inflation. This demonstrates that the theory-driven weights successfully operationalize the construct of educational effectiveness better than unweighted averages.

\begin{figure}[pos=htbp]
  \centering
  \includegraphics[width=0.8\linewidth]{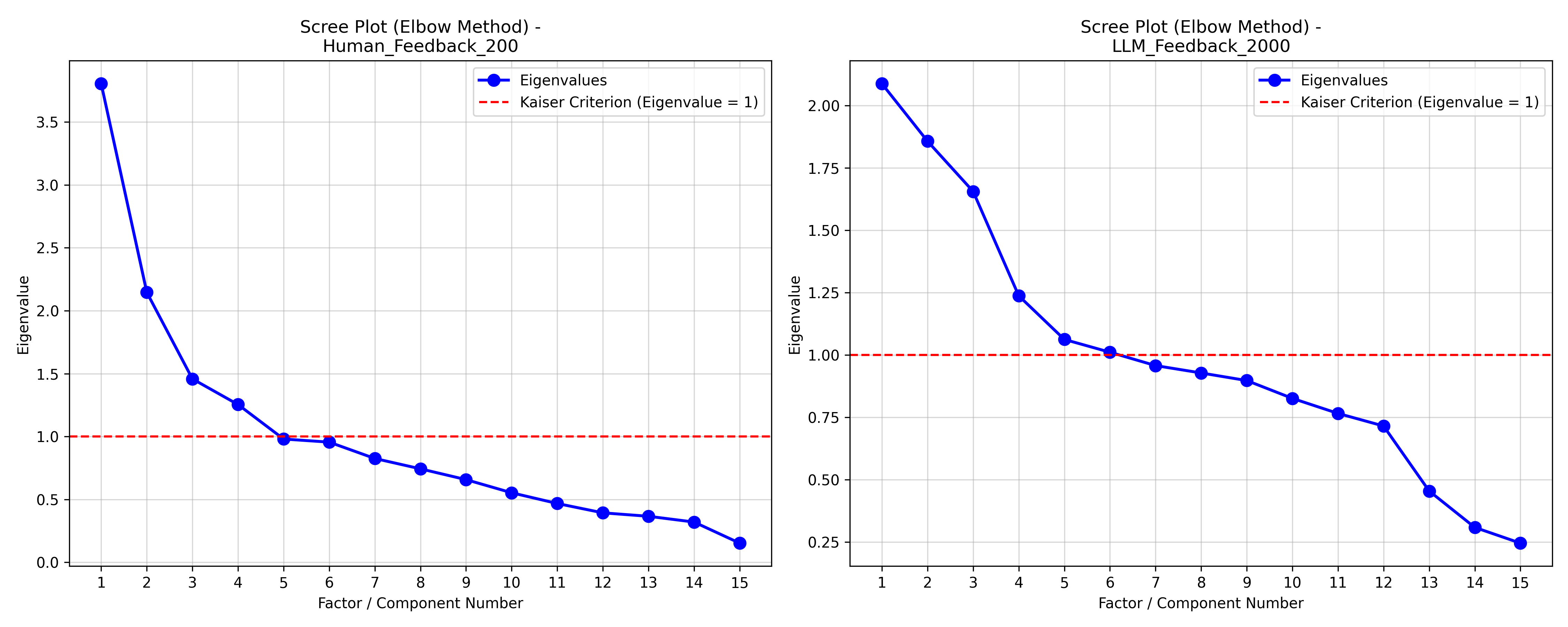}
  \caption{Scree plots used for factor extraction in the human-labelled and LLM-labelled datasets. The Kaiser criterion suggested different factor counts, while the scree-plot elbow supported a five-factor comparison across datasets.}
  \label{fig:combined_scree_elbow}
\end{figure}

The factor-extraction checks supported a five-factor reporting solution for comparing human and LLM feedback-reviewer annotations. The Kaiser criterion suggested four factors for the human-labelled dataset and six factors for the LLM-labelled dataset, whereas the scree plots (Figure \ref{fig:combined_scree_elbow}) showed a visible flattening of eigenvalues after the fifth factor in both datasets. The subsequent exploratory factor analysis therefore used a five-factor solution with Varimax rotation to keep the human and LLM structures comparable. Appendix~\ref{appx:robustness} reports the robustness checks showing that four factors collapsed distinct traits into broader factors, six factors fragmented the LLM-labelled structure, and the five-factor LLM structure remained stable when 20\% of the data were withheld.

The five-factor solution showed alignment on safety and motivational dimensions and divergence on several pedagogical feedback dimensions. Input-Conflicting and Fact-Conflicting hallucinations loaded together in both human and LLM feedback-reviewer annotations, with high absolute loadings in the human-labelled dataset ($>0.89$) and LLM-labelled dataset ($>0.76$). Motivational Tone and Strength also loaded within the same factor in both datasets. By contrast, Feedback on process and Feedback on self-regulation were grouped more closely with other pedagogical indicators in the human-labelled dataset, while the LLM-labelled dataset assigned these dimensions lower maximum absolute loadings ($<0.30$), indicating a more separated rating pattern for these rubric dimensions.

\begin{figure}[pos=htbp]
  \centering
  \includegraphics[width=\linewidth]{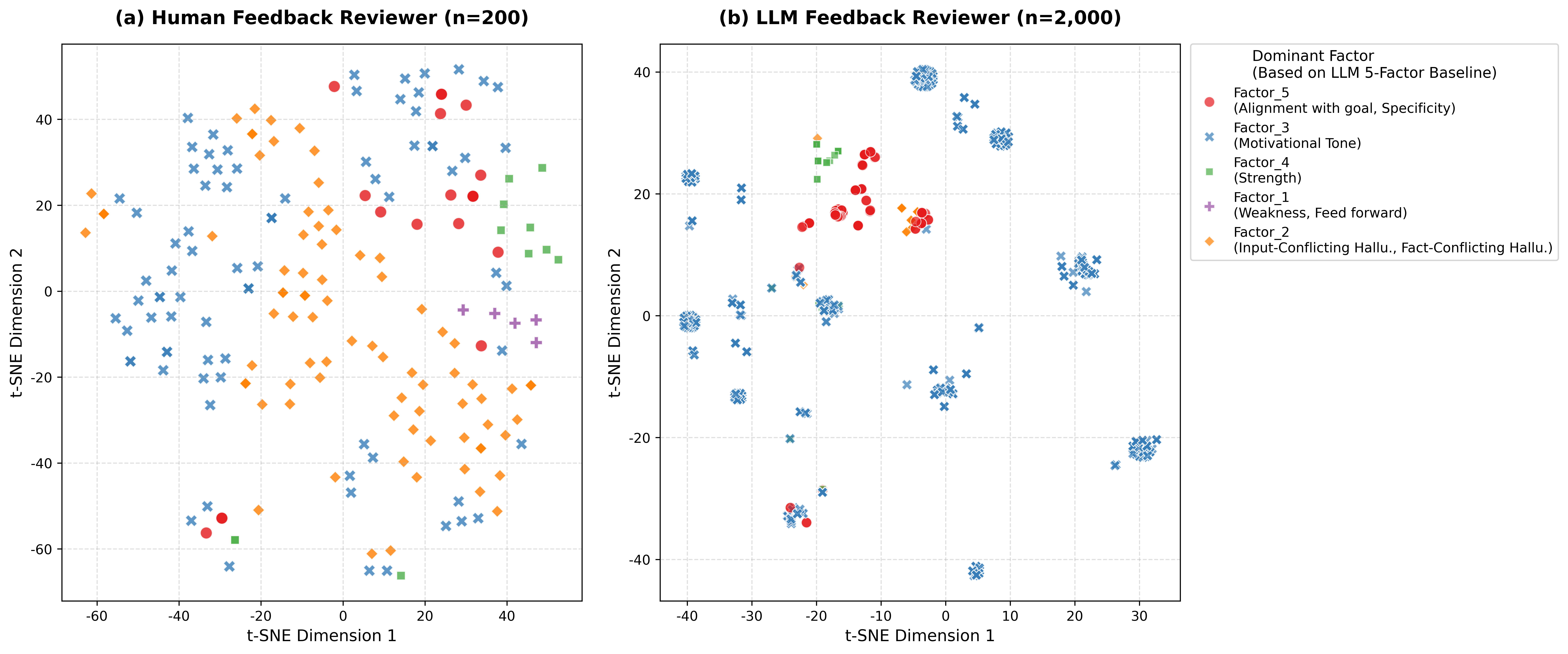}
  \caption{t-SNE visualisation of dominant factor distributions for (a) human feedback-reviewer evaluation results ($n=200$) and (b) LLM feedback-reviewer evaluation results ($n=2{,}000$), using the five-factor LLM structure as the projection basis.}
  \label{fig:Combined_Aligned_TSNE_5Factors}
\end{figure}

The t-SNE projection showed different dominant-factor distributions for human and LLM feedback-reviewer annotations. Using the LLM five-factor framework as the projection basis, LLM-labelled feedback instances were concentrated in Factor 3, corresponding to Motivational Tone ($n=894$). Human-labelled feedback instances were more evenly distributed, with larger groups associated with Motivational Tone ($n=83$) and Safety/Factuality ($n=81$), as shown in Figure~\ref{fig:Combined_Aligned_TSNE_5Factors}.

\FloatBarrier

\subsection{RQ2: Agreement Between LLM Feedback Reviewers and Human Feedback Reviewers}
\label{sec:results_rq2}

RQ2 showed that agreement with the human-labelled ground truth varied by reviewer model, prompting strategy, fine-tuning data, and feedback-quality construct. Table~\ref{tab:summary_deanllms} reports accuracy, Cohen's $\kappa$, and F1 for overall labels and for Content Quality, Educational Effectiveness, and Hallucinations. The table values are reported as agreement and classification indices because confidence intervals were not available in the source output.

\begin{table*}[t]
\caption{Performance of LLM feedback reviewers against the human-labelled ground truth dataset. Acc. = accuracy. Human coder metrics were computed from the inter-rater calibration set.}
\label{tab:summary_deanllms}
\resizebox{\textwidth}{!}{%
\begin{tabular}{ccccccccccccc}
\hline
\multirow{2}{*}{\textbf{\begin{tabular}[c]{@{}c@{}}LLM Feedback\\ Reviewer\end{tabular}}} &
  \multicolumn{3}{c}{\textbf{Overall}} &
  \multicolumn{3}{c}{\textbf{Content Quality}} &
  \multicolumn{3}{c}{\textbf{Educational Effectiveness}} &
  \multicolumn{3}{c}{\textbf{Hallucinations}} \\
 & Acc. & Cohen's $\kappa$ & F1 & Acc. & Cohen's $\kappa$ & F1 & Acc. & Cohen's $\kappa$ & F1 & Acc. & Cohen's $\kappa$ & F1 \\ \hline
\textbf{Zero-shot Prompting} & & & & & & & & & & & & \\
GPT-4.1   & 0.734 & 0.597 & 0.738 & 0.607 & 0.013 & 0.360 & 0.811 & \textbf{0.517} & \textbf{0.756} & 0.807 & 0.115 & 0.519 \\
o4-mini   & 0.738 & 0.607 & 0.742 & \textbf{0.652} & \textbf{0.074} & \textbf{0.391} & 0.780 & 0.464 & 0.726 & 0.813 & 0.145 & 0.536 \\
o3        & 0.741 & 0.603 & 0.742 & 0.603 & 0.012 & 0.358 & 0.816 & 0.483 & 0.741 & 0.833 & 0.300 & 0.632 \\
o3-pro    & \textbf{0.744} & \textbf{0.608} & \textbf{0.745} & 0.607 & 0.008 & 0.366 & \textbf{0.821} & 0.476 & 0.738 & \textbf{0.837} & \textbf{0.308} & \textbf{0.635} \\
Average   & 0.739 & 0.604 & 0.742 & 0.617 & 0.027 & 0.369 & 0.808 & 0.485 & 0.740 & 0.822 & 0.217 & 0.580 \\ \hline
\textbf{Few-shot Prompting} & & & & & & & & & & & & \\
GPT-4.1   & 0.733 & 0.593 & 0.737 & 0.598 & 0.016 & 0.366 & 0.816 & \textbf{0.515} & 0.756 & 0.810 & 0.138 & 0.533 \\
o4-mini   & \textbf{0.749} & \textbf{0.622} & \textbf{0.752} & \textbf{0.657} & \textbf{0.077} & \textbf{0.388} & 0.807 & 0.490 & 0.743 & 0.800 & 0.116 & 0.527 \\
o3        & 0.735 & 0.593 & 0.736 & 0.595 & -0.004 & 0.356 & 0.811 & 0.458 & 0.729 & 0.837 & 0.351 & 0.664 \\
o3-pro    & 0.743 & 0.605 & 0.744 & 0.590 & -0.019 & 0.350 & \textbf{0.824} & 0.489 & 0.745 & \textbf{0.860} & \textbf{0.430} & \textbf{0.703} \\
Average   & 0.740 & 0.603 & 0.742 & 0.610 & 0.017 & 0.365 & 0.815 & 0.488 & 0.743 & 0.827 & 0.259 & 0.607 \\ \hline
\textbf{Fine-tuned GPT-4.1} & & & & & & & & & & & & \\
with plain-labelled data & \textbf{0.798} & \textbf{0.686} & \textbf{0.794} & \textbf{0.737} & \textbf{0.315} & \textbf{0.532} & \textbf{0.843} & \textbf{0.457} & \textbf{0.725} & \textbf{0.813} & \textbf{0.362} & \textbf{0.679} \\
with explanatory data    & 0.721 & 0.566 & 0.705 & 0.675 & 0.000 & 0.269 & 0.799 & 0.332 & 0.667 & 0.633 & 0.237 & 0.590 \\
with both datasets       & 0.703 & 0.532 & 0.674 & 0.675 & 0.000 & 0.269 & 0.810 & 0.377 & 0.687 & 0.507 & 0.175 & 0.499 \\ \hline
\textbf{Human coders} & & & & & & & & & & & & \\
Coder 1   & 0.804 & 0.715 & 0.850 & 0.678 & 0.514 & 0.743 & 0.838 & 0.691 & 0.880 & 0.933 & 0.812 & 0.900 \\
Coder 2   & 0.779 & 0.639 & 0.818 & 0.667 & 0.498 & 0.720 & 0.790 & 0.344 & 0.805 & 0.870 & 0.535 & 0.765 \\
Coder 3   & 0.767 & 0.611 & 0.810 & 0.633 & 0.444 & 0.678 & 0.810 & 0.357 & 0.846 & 0.889 & 0.669 & 0.833 \\
Average   & 0.783 & 0.655 & 0.826 & 0.659 & 0.485 & 0.714 & 0.813 & 0.464 & 0.844 & 0.897 & 0.704 & 0.833 \\ \hline
\end{tabular}%
}
\end{table*}

Prompting-based LLM feedback reviewers showed similar overall agreement across zero-shot and few-shot settings. In the zero-shot condition, o3-pro had the highest overall accuracy (0.744), F1 (0.745), and Cohen's $\kappa$ (0.608). In the few-shot condition, o4-mini had the highest overall accuracy (0.749), F1 (0.752), and Cohen's $\kappa$ (0.622). The average overall values were nearly unchanged from zero-shot to few-shot prompting, with F1 remaining at 0.742 and Cohen's $\kappa$ changing from 0.604 to 0.603.

Agreement patterns differed across the three evaluation constructs. For Educational Effectiveness, o3-pro had the highest accuracy in both zero-shot (0.821) and few-shot (0.824) settings, while GPT-4.1 had the highest Cohen's $\kappa$ in both zero-shot (0.517) and few-shot (0.515) settings. For Hallucinations, o3-pro had the strongest prompting-based performance in both zero-shot (Accuracy = 0.837, $\kappa = 0.308$, F1 = 0.635) and few-shot settings (Accuracy = 0.860, $\kappa = 0.430$, F1 = 0.703). For Content Quality, prompting-based agreement was low across all models, with Cohen's $\kappa$ ranging from -0.019 to 0.077.

Fine-tuned GPT-4.1 showed the highest overall automated reviewer agreement when trained on plain-labelled data. The plain-labelled fine-tuned reviewer achieved overall Accuracy = 0.798, Cohen's $\kappa = 0.686$, and F1 = 0.794, compared with the average human-coder values of Accuracy = 0.783, Cohen's $\kappa = 0.655$, and F1 = 0.826. Plain-labelled fine-tuning also increased Content Quality agreement to Cohen's $\kappa = 0.315$, whereas the prompting-based configurations remained near zero on this construct.

Fine-tuning with explanatory or mixed data showed lower agreement than fine-tuning with plain-labelled data. The explanatory-data condition achieved overall Accuracy = 0.721, Cohen's $\kappa = 0.566$, and F1 = 0.705, while the mixed-data condition achieved overall Accuracy = 0.703, Cohen's $\kappa = 0.532$, and F1 = 0.674. The largest reduction appeared in Hallucinations, where accuracy decreased from 0.813 in the plain-labelled condition to 0.633 in the explanatory-data condition and 0.507 in the mixed-data condition.

\FloatBarrier

\subsection{RQ3: Feedback Quality Across LLM Feedback Tutors}
\label{sec:results_rq3}

\begin{figure}[pos=htbp]
  \centering
  \includegraphics[width=\textwidth]{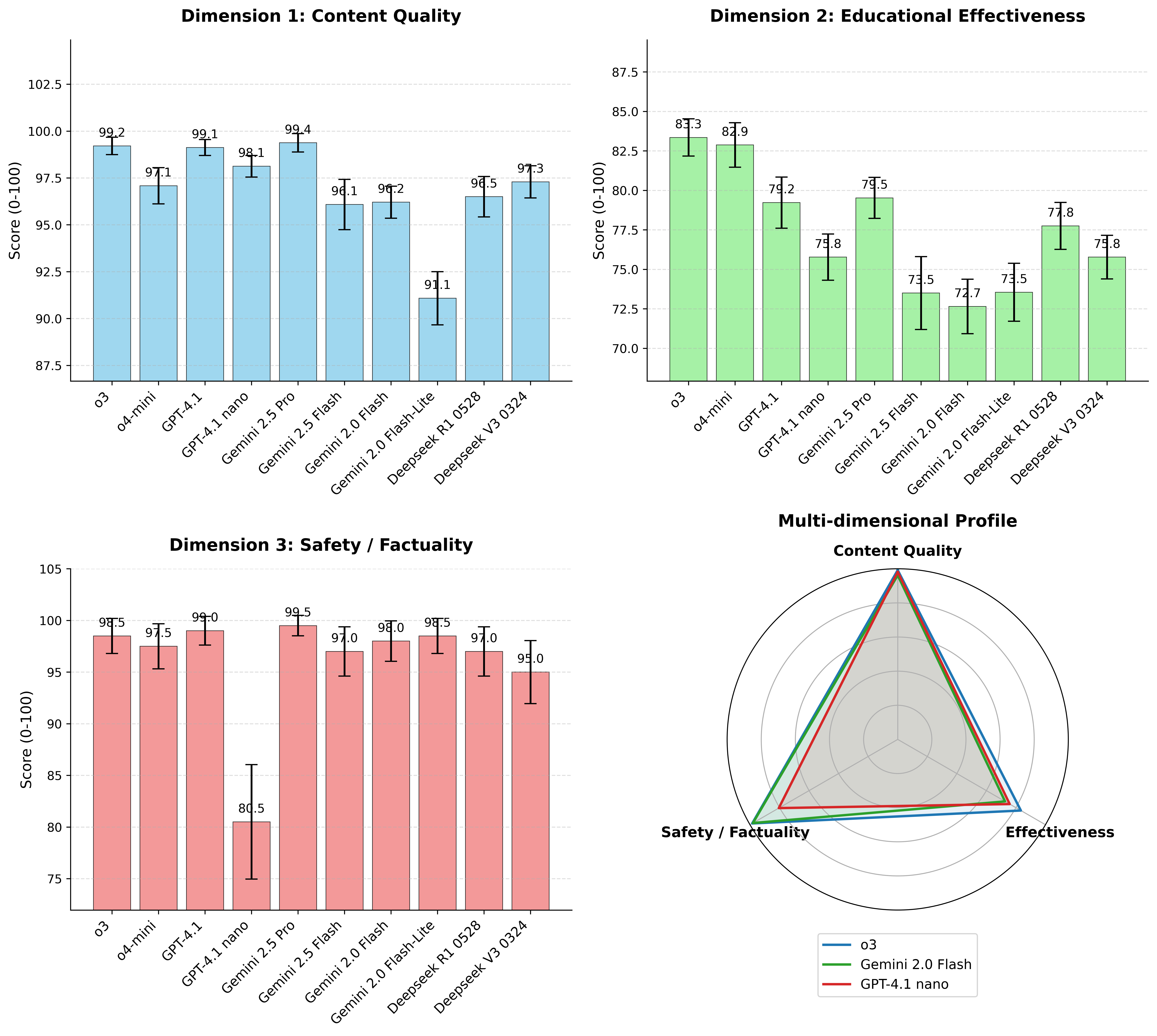}
  \caption{Multi-dimensional performance evaluation of feedback generated by different LLM tutors. The evaluation framework is decomposed into three normalised core dimensions on a 0--100 scale: Content Quality, Educational Effectiveness, and Safety \& Factuality. Error bars in the bar charts denote 95\% confidence intervals.}
  \label{fig:different_llms}
\end{figure}

RQ3 indicated that LLM feedback tutors differed across Content Quality, Educational Effectiveness, and Safety \& Factuality. Figure~\ref{fig:different_llms} reports model-level mean scores on a 0--100 scale, with 95\% confidence intervals shown for the bar charts. The radar chart summarises the same three-dimensional profile for selected representative models.

Content Quality scores were high for most evaluated LLM tutors. Gemini 2.5 Pro (mean = 99.38), o3 (mean = 99.21), and GPT-4.1 (mean = 99.12) had the highest Content Quality scores, and their 95\% confidence intervals overlapped in Figure~\ref{fig:different_llms}. Gemini 2.0 Flash-Lite had the lowest Content Quality score among the evaluated models (mean = 91.08).

Educational Effectiveness showed clearer separation between the highest-scoring tutor models and the remaining models. The OpenAI reasoning models o3 (mean = 83.35) and o4-mini (mean = 82.88) had the highest Educational Effectiveness scores. Gemini Flash models had lower Educational Effectiveness scores, with reported means ranging from 72.65 to 73.55.

Safety \& Factuality scores showed that hallucination-related risk was concentrated in a smaller subset of tutor models. GPT-4.1 nano had the lowest Safety \& Factuality score (mean = 80.50), corresponding to a higher proportion of feedback instances labelled with at least one input-, context-, or fact-conflicting hallucination. Gemini 2.5 Pro (mean = 99.50) and GPT-4.1 (mean = 99.00) had the highest Safety \& Factuality scores among the reported models.

The three RQ3 score dimensions showed different model-ordering patterns rather than a single uniform ranking across all feedback-quality criteria. Models with high Content Quality did not necessarily have the highest Educational Effectiveness, and the Safety \& Factuality score separated GPT-4.1 nano from the models with near-ceiling safety scores. The benchmarking results therefore report the three dimensions separately rather than combining them into a single tutor-quality score.

\section{Discussion}
\label{sec:discussion}

\subsection{Summary of Findings and Research Questions}
\label{sec:discussion_summary}

DeanLLM suggests that automated review of LLM-generated educational feedback is feasible when reviewer calibration is treated as a pedagogical quality-screening problem rather than a generic text-evaluation task. The study developed a 16-dimension framework for evaluating Content Quality, Educational Effectiveness, and Safety \& Factuality; compared human and LLM feedback-reviewer annotations; tested prompting and fine-tuning approaches for reviewer alignment; and benchmarked feedback generated by 10 commercial LLM tutors. Across the three research questions, the findings indicate that LLM feedback reviewers can support scalable pre-delivery evaluation, but that their value depends on construct design, human-labelled calibration, and model selection.

The main contribution of the study is to integrate learner-centred feedback evaluation with hallucination-aware review for LLM tutor feedback. Prior work has established that useful feedback needs to clarify goals, respond to current performance, and support future improvement \citep{h:4,r:3,w:6}, while recent work on LLMs in education has shown that generated feedback may be fluent without being sufficiently accurate, aligned, or pedagogically useful \citep{d:1,k:4,w:7}. DeanLLM addresses this gap by evaluating feedback before it reaches learners, which positions automated review as an additional governance layer for LLM tutors rather than as a replacement for feedback theory, teacher judgement, student uptake, or attention to the emotional consequences of feedback \citep{alsaiari2025emotionally,banihashem2026pedagogical,pozdniakov2026ai}.

\subsection{RQ1: The Evaluation Framework Operates as a Formative Quality Gate}
\label{sec:discussion_rq1}

The psychometric analyses indicate that the 16-dimension framework is better understood as a formative quality gate than as a reflective scale. Human feedback reviewers showed stronger inter-dimensional associations, a structural pattern that aligns with the cognitive ``halo effect'' in human grading \citep{Schmidt2023Halo}, where interrelated feedback traits are judged holistically. In contrast, the LLM feedback reviewer produced a more separated dimension-by-dimension rating profile, demonstrating a strict mechanical orthogonality. This divergence is conceptually significant: while the human halo effect allows educators to integrate tone, specificity, and pedagogical usefulness into a cohesive impression of feedback quality, the automated reviewer isolates these metrics to apply the rubric more mechanically. For DeanLLM, this decoupled pattern is highly advantageous because a pre-delivery review framework needs to preserve strictly separable checks for critical dimensions like hallucination risks, preventing a generally fluent and encouraging text from artificially inflating the scores of distinct safety and factual sub-dimensions.

The construct-level findings support the inclusion of dimensions that are pedagogically necessary even when they show low variance or weak item-total correlations. Clarity and Understandability showed restricted variation, and Context-Conflicting hallucinations appeared sparsely, but both remain relevant to the safe delivery of LLM-generated educational feedback. Learner-centred feedback needs to be understandable and actionable \citep{n:3,s:4,r:3}, and hallucination taxonomies distinguish input-, context-, and fact-related errors because each error type can mislead learners in different ways \citep{z:2,r:2,a:1}. The implication is that automated feedback evaluation should not remove a dimension solely because it is rare or statistically weak in a particular dataset when the dimension represents a meaningful educational risk.

The human--LLM structural divergence also clarifies the role that automated feedback reviewers can play in educational quality assurance. Human reviewers appeared to connect process-level feedback, self-regulation feedback, and broader pedagogical quality more closely, whereas the LLM feedback reviewer separated several advanced pedagogical dimensions from the main factor structure. This separation suggests that LLM feedback reviewers may be well suited for localised rubric checks but less suited for judging the coherence among feedback dimensions unless that relation is explicitly specified. Future automated review frameworks therefore need to distinguish between checking whether a feature is present and evaluating how feedback features work together as a learner-centred feedback artefact.

\subsection{RQ2: Reviewer Calibration Matters More Than Prompt Volume}
\label{sec:discussion_rq2}

The reviewer-agreement results indicate that standard prompting was insufficient for nuanced content-quality judgements. Zero-shot and few-shot reviewers achieved moderate overall agreement in some configurations, and reasoning-oriented models were comparatively stronger for hallucination detection, but prompting-based agreement on Content Quality remained near chance. This pattern is consistent with the nature of feedback evaluation that implies assessing alignment, specificity, and usefulness requires task-grounded interpretation of the assignment, student submission, and feedback text rather than only surface recognition of rubric language \citep{o:1,l:6,s:3,d:4,d:5}. Prompt examples can clarify rubric categories, but they do not necessarily calibrate the reviewer to the local disciplinary and assessment context.

The strongest reviewer performance came from GPT-4.1 fine-tuned with plain-labelled human examples. This finding suggests that compact, task-aligned supervisory signals can improve automated review more effectively than adding longer prompt context or a small number of explanatory rationales. Plain-labelled fine-tuning likely helped the reviewer learn decision boundaries for the specific rubric and dataset, including cases where fluent feedback was still weakly aligned or insufficiently specific. The implication is that educational deployments of LLM feedback reviewers should budget for a small, high-quality calibration set rather than relying only on generic prompting.
The weaker performance of explanatory and mixed fine-tuning should be interpreted as a scope condition rather than evidence against rationale-based supervision. The explanatory dataset was relatively small, and the target outputs combined multi-dimensional labels with open-ended explanations, which may have increased format complexity for an already long evaluation prompt. Prior work suggests that explanations can support model reasoning in some settings, but their value depends on the scale, consistency, and structure of the supervision signal \citep{k:3,b:6,l:11,l:5}. Future studies should compare structured rationales, dimension-specific explanations, and larger rationale datasets before concluding whether explanatory supervision is useful for automated educational feedback review.

\subsection{RQ3: Tutor Model Choice Shapes Feedback Quality and Safety}
\label{sec:discussion_rq3}

The benchmarking results indicate that LLM tutor selection affects the pedagogical quality and factual reliability of generated feedback. Several high-capacity models produced near-ceiling Content Quality scores, but Educational Effectiveness and Safety \& Factuality separated model families more clearly. Reasoning-oriented models showed stronger Educational Effectiveness, which suggests that models with stronger deliberative capabilities may be better able to generate feed-forward, process-level, and self-regulation-oriented feedback. This interpretation is consistent with the broader observation that generating educational feedback requires more than fluent wording; it requires the feedback to connect task evidence, learning goals, and future action \citep{h:4,r:3,d:1,cohn2026theory}.

The lower Safety \& Factuality score for a lightweight model cautions against selecting LLM tutors solely on cost or speed. Lightweight models may be attractive for large-scale educational systems, but a higher hallucination rate is problematic when feedback is delivered as instructional guidance rather than as draft text for teacher revision. Hallucinated feedback can misrepresent student performance, contradict task requirements, or introduce incorrect disciplinary information \citep{z:2,r:2,farquhar2024detecting}. The practical implication is that lower-cost tutor models may require stronger pre-delivery review, stricter safety thresholds, or human-in-the-loop oversight before independent student-facing use, where our DeanLLM architecture may help mitigate limitations of those lower-cost tutor models.

The three-dimensional benchmarking profile shows why feedback quality should not be reduced to a single model ranking. A model can produce well-structured and encouraging feedback while still offering limited process-level guidance or carrying factuality risks. Separating Content Quality, Educational Effectiveness, and Safety \& Factuality makes these trade-offs visible and supports deployment decisions that are aligned with the intended learning context. For example, feedback used for early drafting may prioritise clarity and encouragement, whereas feedback used for summative revision in technical courses requires stronger factuality and task-grounded accuracy.

\subsection{Implications for Educational Research}
\label{sec:discussion_research_implications}

DeanLLM extends automated feedback research by positioning evaluation as a separate layer in LLM tutor systems. Earlier automated feedback evaluation studies have focused on selected dimensions such as feedback level, praise, readability, or correctness \citep{o:1,l:6,s:3,d:4,d:5}. The present framework connects these pedagogical dimensions with hallucination-aware safety checks, which is necessary because LLM-generated educational feedback can be pedagogically weak even when factually correct, and unsafe even when fluent. This integration provides a basis for studying feedback generation and feedback review as coupled but distinct components of educational AI systems.

The findings also show that human and automated feedback review should be studied as different forms of judgement rather than as interchangeable sources of labels. Human reviewers may bring holistic pedagogical interpretation, while LLM feedback reviewers may provide consistent, separable checks across many dimensions. This distinction is important for measurement because high human--LLM agreement is not the only desirable outcome. A reviewer that is more mechanically consistent than a human can be useful for safety gating, but it may miss the relational and pedagogical coherence that human educators consider when judging feedback as a whole.

The study contributes to research on responsible generative AI in education by operationalising pre-delivery review as a measurable design problem. Educational concerns about LLM hallucinations have often been discussed at the level of general risk \citep{y:4,a:1,q:1}, while learner-centred feedback research has emphasised the conditions that make feedback usable for students \citep{h:4,r:3,w:6}. DeanLLM links these traditions by asking whether feedback should be released, revised, or regenerated before students see it. Future research can build on this framing by evaluating reviewer thresholds, missed-risk rates, and the consequences of automated gating for teacher workload and learner trust.

\subsection{Implications for Educational Practice}
\label{sec:discussion_practice_implications}

Educational deployments of LLM tutors should treat automated feedback review as a safeguard rather than as a guarantee of feedback quality. DeanLLM can help institutions screen feedback for alignment, specificity, pedagogical usefulness, and hallucination risk, but the reviewer remains dependent on the rubric, calibration data, and model backbone. A practical deployment should therefore include human-defined quality thresholds, periodic human audits, and clear escalation rules for feedback that fails safety or quality checks.

Teacher-facing systems can use DeanLLM to support iterative feedback generation before student delivery. A tutor model could generate several candidate feedback responses, the reviewer could score them across Content Quality, Educational Effectiveness, and Safety \& Factuality, and the system could either select the strongest candidate or request regeneration when safety thresholds are not met. This workflow preserves learner-centred feedback principles while reducing the likelihood that students receive feedback that is fluent but misaligned, shallow, or factually unreliable.

Model selection for educational feedback should be aligned with the risk profile of the learning task. Reasoning-oriented LLM tutors appear more suitable for feedback that requires process explanation, self-regulation support, and feed-forward guidance, while lightweight models may require stronger oversight if used in student-facing settings. The benchmarking results therefore support a task-sensitive deployment strategy: lower-risk drafting support may tolerate lower-cost models with review, whereas technical, high-stakes, or assessment-adjacent feedback should use stronger tutor and reviewer configurations.

Transparency and learner control should remain part of any system that uses automated feedback review. Students and teachers should know when feedback has been generated by an LLM tutor and reviewed by an automated feedback reviewer, and they should have opportunities to question, revise, or override the feedback when it conflicts with task evidence. Automated review can reduce risk, but responsible deployment still requires consent, accountability, and human oversight because the system evaluates feedback text rather than directly observing student understanding or learning.

\subsection{Limitations and Future Directions} 
\label{sec:discussion_limitations}


Several scope conditions limit how far the findings can be extended beyond automated feedback-quality review. The study used an offline design, so the results concern feedback quality, reviewer agreement, and model benchmarking rather than student uptake, revision behaviour, achievement, engagement, or trust. The synthetic coursework dataset enabled privacy-preserving evaluation and preserved key assignment and feedback-triggering features, but it cannot fully reproduce teacher expectations, learner histories, classroom relationships, multilingual contexts, figures, code execution, or multimodal artefacts. The human-labelled benchmark was also limited in scale, with 200 feedback instances, three expert coders, and a smaller subset of explanatory annotations, which means that difficult dimensions would benefit from larger, adjudicated, and more diverse rating datasets. The reviewer and tutor comparisons are model-specific and time-sensitive because commercial generative AI systems change over time, so future work should repeat the benchmark with explicit model versions, inference settings, closed-source models, and open-weight alternatives. The scoring framework also requires outcome-based validation, particularly for the weighting of Educational Effectiveness and the zero-tolerance Safety \& Factuality rule, because different thresholds may have different consequences for feedback uptake, revision quality, teacher acceptance, and learner trust. The fine-tuning results further suggest the need to compare labels-only supervision with structured rationales, contrastive examples, uncertainty-aware review, and active-learning pipelines focused on cases where human and automated reviewers disagree. Future research should therefore evaluate DeanLLM in live classroom and experimental settings, extend it from single-turn feedback artefacts to multi-turn tutor--learner interaction, and test closed-loop workflows in which the reviewer triggers regeneration, requests evidence, or routes uncertain cases to teachers.

\section{Conclusion}



This study provides evidence that LLM-generated educational feedback could be reviewed systematically using automated methods before it is delivered to learners when pedagogical quality and hallucination risk are evaluated together. DeanLLM operationalised this automated review through a 16-dimension framework covering feedback content, educational effectiveness, and safety-related factuality, and the findings indicate that automated feedback review is best understood as a calibration and governance problem rather than only a text-scoring problem. Human reviewers tended to judge feedback more holistically, whereas the automated reviewer separated rubric dimensions more mechanically, suggesting that LLM feedback reviewers can support scalable quality review but should not be treated as substitutes for educators' contextual judgement. Prompting alone produced limited agreement for nuanced feedback-quality dimensions, while supervised fine-tuning with plain-labelled human examples produced stronger alignment with expert labels, and model benchmarking showed that tutor-model choice also shaped the educational quality and safety of generated feedback. The design implication is that LLM tutors should not be evaluated only by fluency, availability, or generation cost, but also by whether their feedback is aligned with task goals, supports meaningful next steps, and avoids input-, context-, or fact-conflicting hallucinations. DeanLLM therefore points towards an accountable workflow in which automated review can screen feedback, guide regeneration, and support teacher oversight before AI-generated guidance shapes students' learning.

\section{Data and Code Availability}
\label{sec:opensource}

Data and scripts used in the experiments are available in an anonymous Open Science repository: \href{https://anonymous.4open.science/r/DeanLLM-of-LLM-Tutors-E796}{DeanLLM data and code repository}, where you can find the human-labelled feedback quality dataset in data.csv, the 10 LLM Tutor Feedback dataset (1,000 synthetic assignment submissions and corresponding synthetic assignment descriptions) in feedback.csv, the detailed labelling rubric in rubric.docx, and the scripts used in the experiment in code.ipynb.

The LLM tutor feedback reviewer skills are available in a separate anonymous Open Science repository: \href{https://anonymous.4open.science/r/Skill-Dean-of-LLM-Tutors-Anonymous}{DeanLLM reviewer skill repository}, where 4 different LLM tutor feedback reviewer skills used in this paper (introduced in Sec:~\ref{sec:reviewers} and workflow detailed in Appendix~\ref{appx:skill_workflow}) are shared and available for agent usage.

The fine-tuned models can be retrieved by calling OpenAI LLM APIs by theirmodel names.

The GPT-4.1 model fine-tuned on plain-labelled data is ``ft:gpt-4.1-2025-04-14:colam::ByjI44s3''.
The GPT-4.1 model fine-tuned on explanatory data is ``ft:gpt-4.1-2025-04-14:colam::ByjrI6TD''.
The GPT-4.1 model fine-tuned on both datasets is ``ft:gpt-4.1-2025-04-14:colam::BykjbGc5''.

\section{Acknowledgements}
This research was in part supported by [blinded for review] 

\section{Declaration of competing interest}
The authors declare that they have no known competing financial interests or personal relationships that could have appeared to influence the work reported in this paper.

\section{Declaration of generative AI use}

During the preparation of this work, the authors used Gemini, ChatGPT and Claude to improve the grammatical accuracy of the manuscript text and to debug the Python scripts used for statistical evaluation. After using this tool, the authors thoroughly reviewed, verified, and edited the outputs to ensure they accurately reflected the study's findings, and they take full responsibility for the final content.

\appendix

\section{LLM Feedback Reviewer skill}

\label{appx:skill_workflow}

Unlike traditional static systems, LLM agents can dynamically acquire and deploy modular capabilities---commonly referred to as ``skills,'' which are bundles of procedural knowledge, instructions, or code---that extend their capabilities for complex workflows without retraining \citep{Munshi2022Analysing,cohn2026theory}. In education, these skills empower agents to generate instructional content, provide adaptive feedback, simulate classroom interactions, and scaffold both cognitive and non-cognitive competencies \citep{xu2026agent,Wang2025Boosting,Xu2026The}. Empirical studies demonstrate that integrating educational skills into LLM agent workflows can enhance student engagement, metacognitive awareness, writing performance, and problem-solving abilities \citep{xu2026agent,Abdulnour2025Educational}.

\begin{figure}[pos=htbp]
  \centering
    \includegraphics[width=\linewidth]{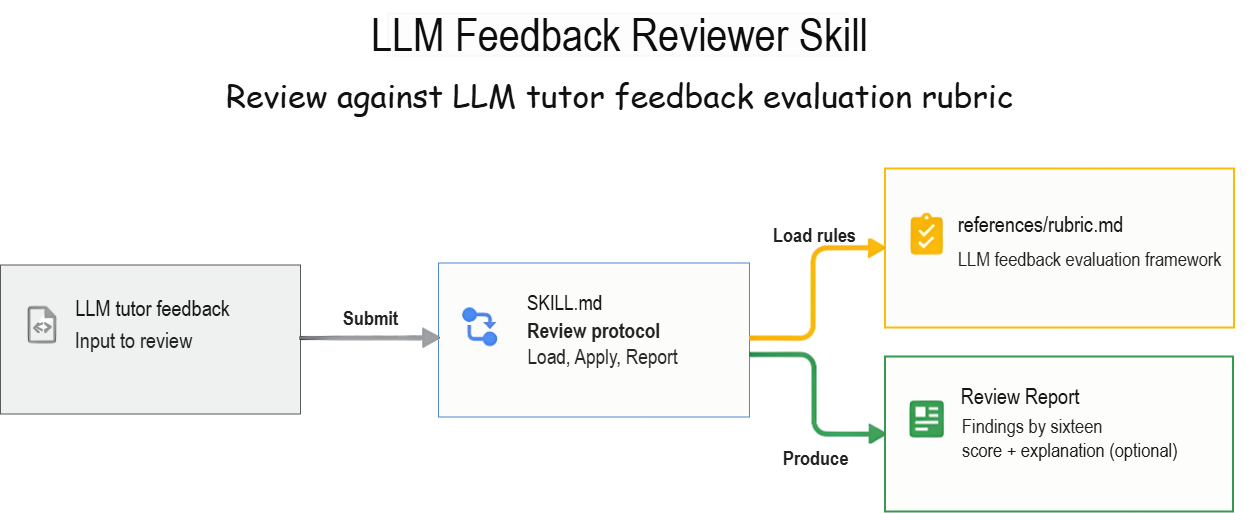}
      \caption{Overview of LLM Feedback Reviewer skill's workflow, adopted from Google's reviewer skill pattern \cite{nigam2026adk}. We have open-source ths LLM feedback reviewer skill, including different settings evaluated in this work.}
      \label{fig:skill}
\end{figure}

To bridge the gap between the demand for scalable automated feedback and the critical need for pedagogical safety, the DeanLLM framework equips an LLM agent with a specialized feedback evaluation skill (Fig.~\ref{fig:skill}) to systematically review and gate tutor-generated feedback before it reaches the student. By introducing an independent, skill-based evaluation layer, this approach ensures that only high-quality, hallucination-free feedback is delivered, advancing the scalable, safe, and effective use of AI in education.

\section{Prompt Example of Zero-shot LLM Feedback Reviewers}
\label{appx:zero-prompt}
\captionof{listing}{Zero-shot prompt for evaluating Alignment with goal dimension.}
\label{lst:zero-shot}
\begin{mdframed}
\begin{lstlisting}
1. Alignment with goals: Alignment with goals evaluates to what extent most comments are aligned to specific goals.
   Score 0: No reference to assignment goals in specific comments.
   Score 1: References goals partly and vaguely.
   Score 2: References goals clearly in
   most comments.
\end{lstlisting}
\end{mdframed}

\section{Prompt Example of Few-shot LLM Feedback Reviewers}
\label{appx:few-prompt}
\captionof{listing}{Few-shot prompt for evaluating Alignment with goal dimension.}
\label{lst:few-shot}
\begin{mdframed}
\begin{lstlisting}
1. Alignment with goal: Alignment with goals evaluates to what extent most comments are aligned to specific learning goals.
   Score 0: No reference to assignment goals in specific comments.
Example Feedback:
"Nice work—keep going!"
(No reference to any assignment goals or criteria.)

   Score 1: References goals partly and vaguely.
Example Feedback:
"Your introduction is okay, but make it stronger to add depth."
(Mentions a goal vaguely but does not tie it to a specific learning outcome or stated goal.)

   Score 2: References goals clearly in most comments.
Example Feedback:
"Excellent work linking your data analysis to the learning outcome of critically evaluating statistical models; each example demonstrates how you tested and interpreted the hypothesis."

"Please also review the FIT citation style as in-text citation is missing in your report."
(Most comments are explicitly framed in objective-language and aligned to the rubric or task requirements or other goals for improvement.)
\end{lstlisting}
\end{mdframed}

\section{Real Assignment Dataset Retrieval}
\label{appx:data_retrieval}

Upon the approval by the [country blinded] University Human Research Ethics Committee (application no. [blinded]), we retrieved the assignment dataset of all university courses delivered during the first semester of 2021 from Faculty of Information Technology of Monash University. We confirm that all methods were performed in accordance with the relevant guidelines and regulations. Within these courses, we collected all assignment materials, including rubrics, reading materials and other assignment supplementary documents, as well as student submissions. The majority of assignment files were provided in the \texttt{.pdf}, \texttt{.txt}, \texttt{.sql}, \texttt{.py}, \texttt{.docx}, \texttt{.ipynb}, and \texttt{.rmd}. We converted them into .txt files, where tables in these files were transformed to structured words that LLMs can understand. This ensured that all assignment files were understandable for LLMs developed since the release of GPT-3.5.

We collected 45,126 student submissions in total, where some of them were too long for some LLMs to take as inputs. We excluded assignments with more than 7,500 words (10,000 tokens for LLMs) in assignment descriptions and corresponding student submissions in total, so that even old and affordable LLMs like GPT-3.5 (with 16k context window) can take assignments in our dataset as input to generate educational feedback to students. We then randomly selected 10,000 student submissions and their corresponding assignment descriptions based, proportionally weighted the selection by the number of submissions per assignment, to reduce cost of LLM-based data synthesis. These 10,000 student submissions spanned 155 assignments across 59 university-level computer science courses.

\section{Data Synthesis}
\label{appx:synthesis}

We used a one-to-one LLM-based imitation method to generate the synthetic assignment descriptions and synthetic student submissions while preserving the original dataset’s semantic meaning and student data distribution, and protecting student privacy and institutional copyright. A real university assignment description was included in the prompt for LLMs to generate a synthetic one. After that, both the real assignment description and the corresponding synthetic one, and a real student submission for the real assignment, were included in the prompt for LLMs to generate a synthetic student submission in response to the synthetic assignment description.

After inclusion of real assignments, we adopted a 4-step mimicry instruction in the prompt:

\begin{enumerate}[label=\arabic*).]
  \item \textbf{Evaluate the original one. } To generate synthetic assignment descriptions, this included theme, objectives, writing style, length and rubric details of the original one. To generate synthetic student assignment, this included i) to what extent, ii) how detailed, and iii) how correctly the real student submission answered the assignment description, where the LLM was instructed to assess the mark of the real submission according to its rubric. The number of words of the real submission was directly added to the prompt.
  \item \textbf{Generate a synthetic one. } The LLM was instructed to imitate the evaluated dimensions in step 1 to create a synthetic assignment description / submission in this step.
  \item \textbf{Evaluate the synthetic one. } The LLM needed to evaluate the generation of step 2 along the same evaluation dimensions as it evaluated the original one.
  \item \textbf{Compare, then output or loop to step 2. } The LLM compared evaluation results of the synthetic one produced in step 3 with the real one produced in step 1. If the LLM thought the results reflecting a successful imitation, then output the synthetic one; else loop to step 2. For example, when generating a synthetic student submission, the LLM needed to compare the mark of the original one concluded in step 1 with the mark of the synthetic one concluded in step 3, together with evaluation results of other dimensions, to make a decision.
\end{enumerate}

Developed from the above framework, details of assignment description synthesis prompts are shown in Appendix.~\ref{appx:description_prompt}, and details of assignment description synthesis prompts are shown in Appendix.~\ref{appx:submission_prompt}.

We proved the similarity between original generated assignments and synthetic assignments in table~\ref{tab:resemblance}.

\begin{table}[!ht]
\caption{Resemblance evaluation results of synthetic assignment descriptions and assignment submissions. BERTScore F1 evalautes sematic similarity. ``PCC'' stands for pearson correlation coefficient of synthetic assignments compared with original real assignments.}
\label{tab:resemblance}
\begin{tabular}{cccc}
\hline
\textbf{Synthetic Data} & \textbf{BERTScore F1} & \textbf{Std} & \textbf{PCC of Length} \\ \hline
Assignment descriptions & 0.859                 & 0.038        & 0.931                    \\
Student submissions     & 0.840                 & 0.020        & 0.852                    \\ \hline
\end{tabular}
\end{table}

We set up an external LLM-based privacy gate to further ensure privacy protection of real students. The complete synthetic student submission generation process was as follows:
\begin{enumerate}[label=\arabic*).]
  \item \textbf{Generate.  } Use the above four-step mimicry framework to generate a synthetic student submission of a synthetic assignment.
  \item \textbf{Judge.  } Use o4-mini (high reasoning effort) to check whether real students' private information was included in the generated synthetic student submission or not. This was done by a three-step instruction prompt: i) Output the student name and id found in the real student submission in the first line; ii) Output the student name and id found in the synthetic one in the second line; iii). If the first line equaled to the second line, output ``YES'' in the third line.
  \item \textbf{Check, then add to the dataset or loop to step 1.  } If ``YES'' was included in step 2 output, then loop to step 1. Else, add step 1 generation results to the synthetic dataset.
\end{enumerate}

We used o3-pro (high reasoning effort) in generation of synthetic assignment descriptions and rubrics, and used o4-mini (high reasoning effort) in generating synthetic student submissions. All models were accessed via API endpoints that ensure the data is not used for future model training, thereby maintaining data privacy, copyright and preventing leakage into future model iterations. We assessed the effectiveness of privacy protection mechanisms using o4-mini (high-reasoning effort) to detect student-identifying information. We found that a perfect privacy protection rate (100\%) was achieved by a combination of our privacy protection mechanisms in the framework that generated our open synthetic dataset.

Details of the data synthesis process and more validity evaluation experiments were disclosed in [Blinded paper for anonymity].

\section{Synthetic Assignment Description Generation Prompt}
\label{appx:description_prompt}

\clearpage
\captionof{listing}{Synthetic assignment description generation prompt format.}
\label{lst:description}
\begin{mdframed}
\begin{lstlisting}
System: You are a helpful university professor. Let's create a university assignment description step by step to be sure you meet all given requirements.

User:
Please try to generate a university assignment description by imitating the theme, objectives, writing style, length and requirement details of the following assignment description.

To do this, let't work this out in a step by step way to be sure we meet all requirements:

Step 1, you need to find out how all questions and goals in the given Assignment descrpition you must imitate;

Step 2, try your best to imitating that the given Assignment descrpition's theme, style and content to generate a virtual assignment description. Remember, you MUST output the complete assignment description no matter how long it will be. You MUST make sure that the assignment description you generate is neither anywhere harder nor easier than the assignment you imitate. You MUST imitating the theme and objectives of the given assignment description, e.g., if the assignment is about generating a database, then you also need to keep the theme and the objectives. You MUST imitating the writing style of the given assignment description. You MUST imitating the length of the given assignment description. You MUST imitating the requirement details of the given assignment description. But you can NOT copy the given assignment description and violating intellectual property.

# Remember, the assignment description you generate MUST keep a similar length as the original one, which is [Insert Number of Words] words. You must check this before you output.
# You MUST only output the generated assignment description.

[[Start of Assignment description you MUST imitate]]

[Insert Original Real Assignment Description]

[[End of Assignment description you MUST imitate]]

# Remember, the assignment description you generate MUST keep a similar length as the original one, which is [Insert Number of Words] words.

Step 3, find out how all questions and
goals in the generated assignment description and check if they equal to the corresponding questions and goals of the given Assignment descrpition you must imitate that you found in step 1. If the assignment you generated is not imitating the original one well, you MUST REGENERATE the assignment description.

Please also count the number of words before you output. If the assignment you generated is either much longer or much shorter than the original one, you MUST REGENERATE the assignment description.

# Be aware
# You MUST only output the generated assignment description, and output the complete assignment description no matter how long it will be.
\end{lstlisting}
\end{mdframed}

\section{Synthetic Assignment Submission Generation Prompt}
\label{appx:submission_prompt}

\captionof{listing}{Synthetic assignment submission generation prompt format.}
\label{lst:generation}
\begin{mdframed}
\begin{lstlisting}
[[Start of Assignment A description]]

[Insert Synthetic Assignment A Description]

[[End of Assignment A description]]

Please act as a virtual student. You need to generate a virtual student submission for Assignment A that has been described above for you. We will also provide Assignment B description and a real student submission for Assignment B to you below, where Assignment A share the same theme, objectives, writing style, length and requirement details as Assignment B.

You job is to try your best to imitating that real student submission's theme, style and content to generate a virtual student submission for Assignment A. You don't need to do anything better or worse than the given student, i.e., you MUST imitate (1). to what extent, (2). how detailed, and (3). how correctly the real student submission answered the assignment description. Thus, your virtual student submission for Assignment B MUST have the same length and the same assignment score as the real student submission for
Assignment A. Remember, you MUST output the complete virtual student submission no matter how long it will be.

To do this, let't work this out in a step by step way to be sure we meet all requirements:

Step 1, you need to find how all questions and goals in Assignment A aligns with questions and goals in Assignment B;

Step 2, evaluate the real student's assignment submission for Assignment B that you need to imitate, see how and to what extent each questions and goals in Assignment B are responsed in the real student assignment submission. Give marks to the submission for each questions and goals in the Assignment B rubric.
This will give you direction on how and to what mark the virtual submission for Assignment A you generated need to response to all questions and goals in the Assignment A rubric that are corresponding to those questions and goals in the Assignment B rubric.

Step 3, try your best to imitating that real student submission's theme, style and content to generate a virtual student submission for Assignment A. You don't need to do anything better or worse than the given student, i.e., you MUST imitate (1). to what extent, (2). how detailed, and (3). how correctly the real student submission answered the assignment description. Thus, your virtual student submission for Assignment B MUST have the same length and the same assignment score as the real student submission for Assignment A. Remember, your virtual submission length MUST be around the length of the real student submission, which is [Insert Number of Words] words.

Step 4, evaluate the virtual student's assignment submission you generated for Assignment A, see how and to what extent each questions and goals in Assignment A are responsed. Give marks to the virtual submission you generated for each questions and goals in the Assignment A rubric, and compare with the corresponding marks for real student submission for Assignment B. If you fonud that the mark of your virtual student assignment is not close to the mark of the given real student assignment, then change the
virtual student assignment you generated until the mark is very close to the mark of the given real student assignment.
\end{lstlisting}
\end{mdframed}

\begin{mdframed}
\begin{lstlisting}
# Be aware

Figures within the Assignment A and B's descriptions and the real student’s submission are excluded from the prompt, so lack of figure is expected.

Tables in the Assignment A and B's descriptions and the real student’s submission are also converted into plain text, in a way which you are able to detect and understand.

The Assignment A and B may require multiple submission files, but some required files may not be included in the real student's submission. The provided student's assignment submission is either the single-submission file or one of the submission files of the student. Thus, you don't need to complement things in your submission if some requried submission are missing from the student assignment submission. You ONLY need to imitate the given student submission as required.

# Be aware

REMEMBER, you ONLY need to output the virtual student's submission to the Assignment A; you MUST not include any of the real student's identity (e.g., name, student ID) and other personal information; you must protect the real student's privacy.

[[Start of Assignment B description]]

[Insert Original Real Assignment B description]

[[End of Assignment B description]]


[[Start of the real student's assignment submission for Assignment B that you need to imitate to answer similar assignment Assignment A]]

[Insert Real Student's Assignment Submission for Assignment B]

[[End of the real student's assignment submission for Assignment B that you need to imitate to answer similar assignment Assignment A]]

REMEMBER, you ONLY need to output the virtual student's submission to
the Assignment A, and output the complete virtual student's submission no matter how long it will be..'''
\end{lstlisting}
\end{mdframed}

\newpage

\section{Evaluation of LLM Tutor Feedback Quality for Real and Synthetic Assignments}
\label{appx:feedback_evaluation}

Results are shown in Table~\ref{tab:feedback_similarity}. More validity evaluation experiments were disclosed in [Blinded paper for anonymity].

\begin{table}[pos=htbp]
\caption{Metrics comparing LLM feedback quality for real assignments and corresponding synthetic assignments. Alignment denotes the percentage of agreement between attributes of LLM tutor feedback for synthetic submissions and those generated for their original real-world submissions. Sensitivity, Specificity, and F1 Score are calculated treating feedback attributes of LLM tutor feedback for original real-world submissions as the ground truth.}
\label{tab:feedback_similarity}
\centering
\resizebox{\linewidth}{!}{%
\begin{tabular}{lllll}
\toprule
\textbf{Attribute} & \textbf{Alignment (95\% CI)} & \textbf{Sensitivity (95\% CI)} & \textbf{Specificity (95\% CI)} & \textbf{F1 Score (95\% CI)} \\
\midrule
Upcoming Similar Tasks & 0.983 (0.974, 0.990) & 0.989 (0.983, 0.995) & 0.000 (0.000, 0.000) & 0.991 (0.987, 0.995) \\
Strengths and Weaknesses & 0.998 (0.995, 1.000) & 1.000 (1.000, 1.000) & 0.333 (0.000, 1.000) & 0.999 (0.997, 1.000) \\
Performance Summary & 0.774 (0.748, 0.799) & 0.887 (0.866, 0.908) & 0.304 (0.243, 0.376) & 0.864 (0.846, 0.880) \\
Meeting Learning Objective & 0.999 (0.997, 1.000) & 0.000 (0.000, 0.000) & 0.999 (0.997, 1.000) & 0.000 (0.000, 0.000) \\
Active Role & 0.695 (0.669, 0.724) & 0.471 (0.421, 0.527) & 0.825 (0.796, 0.855) & 0.531 (0.483, 0.577) \\
Affirmation and Encouragement & 0.707 (0.678, 0.736) & 0.571 (0.512, 0.628) & 0.756 (0.726, 0.788) & 0.509 (0.461, 0.557) \\
Student Teacher Relationship & 0.771 (0.743, 0.796) & 0.576 (0.520, 0.629) & 0.859 (0.830, 0.886) & 0.610 (0.564, 0.658) \\
\textbf{Average} & \textbf{0.847 (0.838, 0.855)} & \textbf{0.855 (0.843, 0.866)} & \textbf{0.837 (0.825, 0.849)} & \textbf{0.856 (0.848, 0.865)} \\
\bottomrule
\end{tabular}
}
\end{table}

\section{LLM Tutor Feedback Generation Prompt}
\label{appx:tutor_prompt}


\captionof{listing}{LLM tutor feedback generation prompt format.}
\label{lst:tutor}
\begin{mdframed}
\begin{lstlisting}
[[Start of assignment description]]
(insert assignment description here)
[[End of assignment description]]

[[Start of the student's assignment submission]]
(insert the student's assignment submission here)
[[End of the student's assignment submission]]

Please act as a virtual tutor. Your goal is to generate educational feedback for the given student’s assignment submission of the given assignment.
# Be aware
Figures within the assignment description and the student’s submission are excluded from the prompt, so lack of figure is not the student’s fault.
Tables in the assignment description and student’s submission are also converted into plain text, in a way which you are able to detect and understand.
The assignment may require multiple submission files, but some required files may not be included here. The provided student's assignment submission is either the single-submission file or one of the submission files of the student. The student's real personal information (name, student ID) are also removed or changed from the submission. Thus, you don't need to comment on the missing parts/files of the student submission.

Finally, your feedback MUST be less than 300 words.
\end{lstlisting}
\end{mdframed}
\clearpage

\section{Sensitivity Analysis of Heuristic Composite Scoring}
\label{appx:sensitivity}

To empirically validate the robust nature of the composite scoring formulas and veto rules established in Method Sec.~\ref{sec:feedback_scores}, a comprehensive sensitivity analysis was performed. We tested two primary systemic variations against the baseline metrics: (1) replacing the theory-driven weights for Educational Effectiveness with an unweighted equal-distribution scheme across all sub-dimensions, and (2) implementing a graduated soft penalty for Safety \& Factuality (averaging errors) rather than the strict zero-tolerance binary gate. Table~\ref{tab:sensitivity_results} outlines the structural shifts in performance across all ten commercial LLM tutors under these alternative mathematical assumptions.

\begin{table}[pos=htbp]
\centering
\caption{Sensitivity Analysis Comparing Baseline and Alternative Scoring Configurations ($n=2,000$).}
\label{tab:sensitivity_results}
\resizebox{\textwidth}{!}{
\begin{tabular}{lcccc}
\toprule
\textbf{Model Group} & \textbf{Effectiveness (Baseline)} & \textbf{Effectiveness (Equal-Weight)} & \textbf{Safety (Baseline Gate)} & \textbf{Safety (Graduated Soft)} \\
\midrule
o3 & 83.35 & 79.50 & 98.50 & 99.33 \\
o4-mini & 82.88 & 81.79 & 97.50 & 98.83 \\
GPT-4.1 & 79.22 & 80.12 & 99.00 & 99.33 \\
GPT-4.1 nano & 75.78 & 74.33 & 80.50 & 89.17 \\
Gemini 2.5 Pro & 79.53 & 81.33 & 99.50 & 99.83 \\
Gemini 2.5 Flash & 73.50 & 77.08 & 97.00 & 98.00 \\
Gemini 2.0 Flash & 72.65 & 69.94 & 98.00 & 98.83 \\
Gemini 2.0 Flash-Lite & 73.55 & 70.00 & 98.50 & 99.50 \\
Deepseek R1 0528 & 77.75 & 74.69 & 97.00 & 98.33 \\
Deepseek V3 0324 & 75.78 & 75.83 & 95.00 & 97.17 \\
\bottomrule
\end{tabular}
}
\end{table}

The empirical results of the sensitivity analysis provide critical validation for the core framework architecture. First, the comparison within the Educational Effectiveness construct demonstrates that an unweighted equal scheme artificially compresses the differentiation between high-capacity reasoning models and lighter alternatives. For instance, the pedagogical depth of the reasoning-optimized o3 model drops visibly from 83.35 to 79.50 when its advanced process-level and self-regulation feedback indicators are not assigned higher priority. Concurrently, models characterized by higher volumes of generalized surface praise receive unearned score inflation under equal weights. This behavior directly confirms that the theoretical coefficients are necessary to capture authentic cognitive scaffolding.

Second, the contrast between the safety indicators underscores the mathematical necessity of the strict zero-tolerance veto mechanism. Under the graduated soft penalty configuration, the severe safety risks present in lightweight models are obscured; for example, GPT-4.1 nano experiences a deceptive score inflation from an unacceptable 80.50 up to a seemingly viable 89.17. In student-facing deployments, a single unvetted contextual or factual hallucination carries the potential to fundamentally misguide a learner. A smoothed average configuration dilutes this critical signal, whereas the baseline strict maximum-penalty framework successfully surfaces localized vulnerabilities. Consequently, the scoring design utilized in the main pipeline stands validated as an indispensable pedagogical safeguard.

\section{Sensitivity and Robustness Evaluations for EFA}
\label{appx:robustness}

To ensure the methodological rigor and replicability of our structural findings regarding the LLM reviewer annotation result paradigm, we conducted comprehensive sensitivity and robustness checks on the extracted exploratory factor models using \textit{scikit-learn}. The evaluations tested forced factor constraints, loading thresholds, and sub-sampling stability.

\subsection{Sensitivity to Forced Factor Extraction Constraints}

To rigorously justify the extraction of 5 latent factors across both populations (superseding the Kaiser criterion), we tested alternative baseline models extracting 4 and 6 factors for the LLM reviewer annotation dataset. We evaluated model stability based on the occurrence of cross-loadings (items loading $>0.30$ on more than one factor) and unassigned items (items failing to reach $>0.30$ on any factor). The detailed item distributions are presented in Table~\ref{tab:forced_factors}.

\begin{table}[h]
\caption{Sensitivity Analysis of Forced Factor Extraction on the LLM Dataset ($n=2,000$)}
\label{tab:forced_factors}
\centering
\renewcommand{\arraystretch}{1.3}
\begin{tabular}{p{0.15\linewidth} p{0.25\linewidth} p{0.5\linewidth}}
\hline
\textbf{Extracted Factors ($k$)} & \textbf{Cross-loadings ($>0.30$)} & \textbf{Unassigned Items ($<0.30$ on all factors)} \\
\hline
$k=4$ & 0 & 6 items: \textit{Feed back, Feedback on process, Feedback on self-regulation, Feedback on self, Clarity and Understandability, Context-Conflicting Hallucinations} \\
$k=5$ (Baseline) & 0 & 5 items: \textit{Feedback on process, Feedback on self-regulation, Feedback on self, Clarity and Understandability, Context-Conflicting Hallucinations} \\
$k=6$ & 1 item: \textit{Strength} & 6 items: \textit{Feed up, Feedback on process, Feedback on self-regulation, Feedback on self, Clarity and Understandability, Context-Conflicting Hallucinations} \\
\hline
\end{tabular}
\end{table}

As shown in Table~\ref{tab:forced_factors}, extracting 4 factors artificially compressed the latent space, leaving 6 critical instructional metrics unassigned. Conversely, extracting 6 factors resulted in structural fragmentation, inducing a cross-loading for the \textit{Strength} metric and causing \textit{Feed up} to drop out of the primary factor structure. The 5-factor constraint serves as the mathematical optimum, achieving zero cross-loadings while minimizing the number of unassigned peripheral metrics, thereby preventing both over-extraction and artificial blurring of latent constructs.

\subsection{Robustness to Factor Loading Thresholds}

To verify the stability of the retained measurement items, we subjected the factor loadings in our 5-factor baseline model to increasingly stringent inclusion thresholds ($>0.40$ and $>0.50$, compared to the standard $>0.30$). The retention rates and specific discarded metrics are summarized in Table~\ref{tab:loading_thresholds}.

\begin{table}[h]
\caption{Retention of Core Metrics Under Varying Loading Thresholds}
\label{tab:loading_thresholds}
\centering
\renewcommand{\arraystretch}{1.3}
\begin{tabular}{p{0.15\linewidth} p{0.2\linewidth} p{0.55\linewidth}}
\hline
\textbf{Threshold} & \textbf{Retained Items} & \textbf{Discarded Edge Metrics (Max Loading $\le$ Threshold)} \\
\hline
$>0.30$ (Standard) & 10 / 15 (66.7\%) & \textit{Feedback on process, Feedback on self-regulation, Feedback on self, Clarity, Context-Conflicting Hallu.} \\
$>0.40$ (Strict) & 8 / 15 (53.3\%) & Above + \textit{Specificity, Feed up} \\
$>0.50$ (Rigorous) & 7 / 15 (46.7\%) & Above + \textit{Alignment with goal} \\
\hline
\end{tabular}
\end{table}

Under the standard $0.30$ threshold, 10 out of 15 core metrics were safely retained. Even under the most rigorous $>0.50$ criterion, 7 core metrics driving Content Quality and Factuality safely retained their dominant factor assignments with minimal degradation. Items discarded under higher thresholds were predominantly those pre-assigned a theoretical weight of 0.0 (e.g., \textit{Feedback on self}) or highly contextual metrics operating under a maximum-penalty mechanism (e.g., \textit{Context-Conflicting Hallucinations}). This corroborates that the core evaluation framework rests on highly resilient statistical foundations.

\subsection{Sub-sampling Stability and Outlier Resistance}

To ensure that the observed extreme orthogonality in LLM reviewer annotation results was not a statistical artifact driven by a specific subset of outlier responses, we conducted a rigorous sub-sampling stability check. We randomly withheld 20\% of the LLM dataset ($n=400$, random state=999) and re-estimated the 5-factor baseline model on the remaining 80\% subset ($n=1,600$). We then evaluated the stability of the latent structure by directly comparing the maximum absolute factor loadings of the core retained metrics across both models.

The empirical results of this split-sample validation are detailed in Table~\ref{tab:sub_sampling}.

\begin{table}[h]
\caption{Comparison of Maximum Factor Loadings Between the Full Dataset ($n=2,000$) and an 80\% Random Sub-sample ($n=1,600$).}
\label{tab:sub_sampling}
\centering
\renewcommand{\arraystretch}{1.3}
\begin{tabular}{l c c c}
\hline
\textbf{Core Measurement Item} & \textbf{Loading (Full 100\%)} & \textbf{Loading (Subset 80\%)} & \textbf{$\Delta$ (Absolute Diff)} \\
\hline
Alignment with goal                 & 0.496 & 0.497 & 0.000 \\
Specificity                         & 0.388 & 0.407 & 0.019 \\
Motivational Tone                   & 0.947 & 0.948 & 0.001 \\
Strength                            & 0.522 & 0.526 & 0.004 \\
Weakness                            & 0.796 & 0.777 & 0.020 \\
Feed forward                        & 0.859 & 0.899 & 0.040 \\
Feed up                             & 0.350 & 0.340 & 0.010 \\
Feed back                           & 0.542 & 0.549 & 0.006 \\
Input-Conflicting Hallu.            & 0.976 & 0.980 & 0.004 \\
Fact-Conflicting Hallu.             & 0.761 & 0.749 & 0.012 \\
\hline
\multicolumn{3}{l}{\textit{Mean Absolute Difference across all retained items:}} & \textbf{0.012} \\
\hline
\end{tabular}
\end{table}

The alignment between the full-scale model and the sub-sampled configuration demonstrated robust stability. Specifically, 100\% of the 10 core metrics that cleared the standard $>0.30$ threshold in the full model successfully loaded onto identical latent factors in the 80\% subset. The fluctuations in their loading magnitudes were negligible, yielding a remarkably low Mean Absolute Difference ($\Delta$) of just 0.012. Even the maximum observed divergence, present in the \textit{Feed forward} metric, was confined to a minor variance of 0.040. No cross-loading shifts or dimensional structural collapse occurred under the sample reduction.

This cross-validation empirically confirms that the mechanical, decoupled generation strategy of the LLM is a systemic and robust phenomenon inherent to the model's behavior, rather than an artifact driven by localized data noise or outlier instances. The resilience of these factor loadings provides solid methodological support for utilizing this 5-factor structure as a baseline for cross-population evaluations.

\section{Example of Human Feedback Annotation (with Explanation)}
\label{appx:annotation_example}

\clearpage
\captionof{listing}{An Example of Human Feedback Reviewer Annotation (with Explanation), and the Corresponding LLM Tutor Feedback.}
\label{lst:annotation_example}
\begin{mdframed}
\begin{lstlisting}[basicstyle=\footnotesize\ttfamily, breaklines=true]
Feedback annotation example 

Alignment with goal
2
Specificity
2
Motivational Tone
2
Strength
2
Weakness
2
Feed forward
1
Feed up
1
Feed back
1
Feedback on task
1
Feedback on process
1
Feedback on self-regulation
1
Feedback on self
1
Clarity and Understandability
2
Input-Conflicting Hallucinations
1
Context-Conflicting Hallucinations
1
Fact-Conflicting Hallucinations
0
\end{lstlisting}
\end{mdframed}

\begin{mdframed}
\begin{lstlisting}[basicstyle=\footnotesize\ttfamily, breaklines=true]
Explanation (optional)
This feedback is Specificity 2 as all comments offer detailed, actionable guidance, with some comment(s) including concrete examples of how to improve, like 'While more is not penalized, you could clarify in your comments that you are exceeding the minimum' where it offers an example 'you are exceeding the minimum' about how to clarify in comments.
'While more is not penalized, you could clarify in your comments that you are exceeding the minimum.' is Feed forward.
'For clarity, you could group the wild and bred animal inserts separately, with section comments.' is Feed forward.
'Consider adding a brief comment for each breeding event, indicating which animals are the parents and which are the offspring.' is Feed forward.
'''
- You inserted 12 animals, with a mix of wild-caught (brevent_id NULL) and bred animals (with brevent_id), as required.
- You created 4 breeding events, each with valid parent animal IDs, and then inserted offspring referencing these events.
- All animal_id and brevent_id values are hardcoded and below 100, as specified.
- Dates for breeding events are within the allowed range (01-Jan-2020 to 31-Jan-2021).
- You correctly avoided placing any Tasmanian Devil animals in AUS10 or AUS20, following the special rule.
- The script is readable, with comments indicating which animals are wild and which are bred.
''' are Feed back.
'You have met all the requirements for this task' is Feed up.
'Your comments at the top are helpful and show awareness of the assignment requirements.' is Feed up.
'''
You have provided a clear and well-structured SQL script for inserting test data into the ANIMAL and BREEDING_EVENT tables. Your comments at the top are helpful and show awareness of the assignment requirements.
While more is not penalized, you could clarify in your comments that you are exceeding the minimum.
you could group the wild and bred animal inserts separately, with section comments.
Consider adding a brief comment for each breeding event, indicating which animals are the parents and which are the offspring, to make the relationships even clearer for the marker.
You have met all the requirements for this task, and your SQL is correct and well-documented.
''' are Feedback on task.
'''
you could clarify in your comments that you are exceeding the minimum.
you could group the wild and bred animal inserts separately
Consider adding a brief comment for each breeding event
''' are Feedback on process.
'Consider adding' is Feedback on self-regulation.
'Great job!' is Feedback on self.
'''
\end{lstlisting}
\end{mdframed}

\begin{mdframed}
\begin{lstlisting}[basicstyle=\footnotesize\ttfamily, breaklines=true]
For Hallucinations:
The feedback first says ``You inserted 12 animals, with a mix of wild-caught … and bred animals,'' implying the script contains only 12 animals.  
However, the student script actually inserts 20 animals (12 wild + 8 bred).  
Later, in the ``Areas for Improvement'' section, the feedback acknowledges ``12 wild animals, then 8 bred animals, for a total of 20 animals.''  
Thus:  

• Input-conflicting: the initial statement that only 12 animals were inserted contradicts the student’s submission.  
• Context-conflicting: the feedback itself contains two mutually contradictory counts (12 animals vs 20 animals).  

No statements conflict with general real-world facts, so Fact-conflicting hallucinations are not present.
\end{lstlisting}
\end{mdframed}

\begin{mdframed}
\begin{lstlisting}[basicstyle=\footnotesize\ttfamily, breaklines=true]
Corresponding LLM tutor feedback:

**Feedback for T2-cb-insert.sql**

Hi Dea,

You have provided a clear and well-structured SQL script for inserting test data into the ANIMAL and BREEDING_EVENT tables. Your comments at the top are helpful and show awareness of the assignment requirements.

**Strengths:**
- You inserted 12 animals, with a mix of wild-caught (brevent_id NULL) and bred animals (with brevent_id), as required.
- You created 4 breeding events, each with valid parent animal IDs, and then inserted offspring referencing these events.
- All animal_id and brevent_id values are hardcoded and below 100, as specified.
- Dates for breeding events are within the allowed range (01-Jan-2020 to 31-Jan-2021).
- You correctly avoided placing any Tasmanian Devil animals in AUS10 or AUS20, following the special rule.
- The script is readable, with comments indicating which animals are wild and which are bred.

**Areas for Improvement:**
- You inserted 12 wild animals, then 8 bred animals, for a total of 20 animals. The requirement was a minimum of 12 animals in total, with some bred and some wild. While more is not penalized, you could clarify in your comments that you are exceeding the minimum.
- For clarity, you could group the wild and bred animal inserts separately, with section comments.
- Consider adding a brief comment for each breeding event, indicating which animals are the parents and which are the offspring, to make the relationships even clearer for the marker.

**Overall:**  
You have met all the requirements for this task, and your SQL is correct and well-documented. Great job!

**Mark: 25/25**
\end{lstlisting}
\end{mdframed}

\bibliographystyle{cas-model2-names}
\bibliography{cas-refs}

\end{document}